\documentclass[reprint,nofootinbib]{revtex4-1}

\usepackage[margin=1in]{geometry} 
\usepackage{amsmath}
\usepackage{amsthm}
\usepackage{amssymb}
\usepackage{tikz}
\usetikzlibrary{matrix,arrows,shapes.geometric,backgrounds}
\newcommand{\e}[1]{{\mathbb E}}
\newcommand{\mymatrix}{
    \left(\begin{gathered}
    \tikzpicture[every node/.style={anchor=south west}]
        \node[minimum width=1cm,minimum height=2cm] at (0,0) {$q_{12}$};
        \node[minimum width=2cm,minimum height=1cm] at (1,2) {$q_{12}$};
        \node[minimum width=1cm,minimum height=1cm] at (0,2) {$q_{11}$};
        \node[minimum width=2cm,minimum height=2cm] at (1,0) {$q_{22}$};

        \draw[dashed] (1,0) -- (1,3);
        \draw[dashed] (0,2) -- (3,2);
    \endtikzpicture
    \end{gathered}\right)
}
\usepackage[utf8]{inputenc}
\usepackage{hyperref}
\hypersetup{
	unicode,
	pdfauthor={Author One, Author Two, Author Three},
	pdfproducer={LaTeX},
	pdfcreator={pdflatex}
}

\usepackage{graphicx, color}
\begin{document}
\title{The dynamics of opinion expression}
\author{Felix Gaisbauer, Eckehard Olbrich, and Sven Banisch}

\affiliation{%
 Max Planck Institute for Mathematics in the Sciences\\
 Inselstrasse 22, 04103 Leipzig
}%


\begin{abstract}
Modelling efforts in opinion dynamics have to a large extent ignored that opinion exchange between individuals can also have an effect on how willing they are to express their opinion publicly. Here, we introduce a model of public opinion expression. Two groups of agents with different opinion on an issue interact with each other, changing the willingness to express their opinion according to whether they perceive themselves as part of the majority or minority opinion. We formulate the model as a multi-group majority game and investigate the Nash equilibria. We also provide a dynamical systems perspective: Using the reinforcement learning algorithm of $Q$-learning, we reduce the $N$-agent system in a mean-field approach to two dimensions which represent the two opinion groups. This two-dimensional system is analyzed in a comprehensive bifurcation analysis of its parameters. The model identifies social-structural conditions for public opinion predominance of different groups. Among other findings, we show under which circumstances a minority can dominate public discourse.
\end{abstract}

\maketitle
\begin{quote}
``The actual strength of [...] different camps of opinion does not necessarily determine which view will predominate in public. An opinion can dominate in public and give rise to the pressure of isolation even if the majority of the population holds the opposing view that has come under pressure---yet does not publicly admit to holding this position."\cite{noelle2004spiral}
\end{quote}

\section{Introduction}

Fundamental to models of opinion dynamics is the assumption that people's opinions are, in some way or another, influenced by the opinion of their peers. There is an extensive amount of models of opinion change in social systems (see \cite{LORENZ2007,Castellano2009, flache2017} for reviews). While it is a plausible assumption that people who express their opinion about an issue are sensitive to approval and disapproval, feedback on the opinion need not necessarily lead to its reconsideration. It might also affect one's willingness of opinion \textit{expression}: The more positive (negative) the feedback, the more (less) motivated one feels to publicly express one's opinion.

In comparison, this approach to public discourse has remained, from a modelling perspective, rather unexplored. However, it is worth to be considered: In general, people are not always willing to reveal their opinion on certain issues to others \cite{matthes}. A recent study shows that only a minority of users who consume news online is also involved in sharing and discussing them \cite{kalogeropulos}. Thorough research on opinion dynamics must take into account that some individuals might choose to not express their opinion publicly, which has profound effect on how others perceive the opinion climate in a social system. We will hence, in this paper, focus on a model of the \textit{expression of}, and not the change in, opinions.

An theory of public opinion expression has already been developed around fifty years ago, with Elisabeth Noelle-Neumann's influential `spiral of silence'  \cite{NoelleNeumann1974,noelle2004spiral}. Roughly speaking, Noelle-Neumann sees the fear of isolation as an essential drive for how humans publicly behave. Especially with respect to morally charged topics, individuals constantly and mostly sub-consciously monitor the `opinion landscape' around them (they possess a ``quasi-statistical sense'' \cite{NoelleNeumann1974,noelle2004spiral}) and might refrain from expressing their opinion if they believe to be part of the minority. On the other hand, a belief to hold the majority position might encourage them to express their view. Since each individual's decision whether to express her opinion or not influences how others perceive the opinion landscape, whose evaluation might then change accordingly, a dynamical development (for which Noelle-Neumann used the metaphor of a spiral) follows in which the seemingly dominant opinion fraction becomes more and more vocal and the perceived minority fraction becomes more and more silent. Noelle-Neumann's spiral of silence is particularly interesting for mathematical modelling since it links a micro mechanism with a dynamical development at the macro level.

While there have been attempts to model opinion expression and specifically the spiral of silence, they are either in large parts simulative \cite{Sohngeidner,Ross,Takeuchi2015,annie2011simulation,gawronski2014opinion} or directed towards the effect of specific circumstances on the spiral of silence (mass media \cite{Sohn2019}, social bots \cite{Ross}, or the long-time effect of charismatic agents \cite{gawronski2014opinion}). Granovetter and Soong \cite{GranovetterSoS}, and subsequently Krassa \cite{Krassa1988}, employ a threshold model of opinion expression which only applies to cases in which a certain opinion is already suppressed. We aim here towards a more general, structural understanding of the dynamics of opinion expression.

We develop a model which employs an account of social influence termed social feedback theory \cite{sft}. The behavioral adjustment of agents depends solely on the social feedback they receive when they express their opinion. This affective experience-based interaction mechanism has already been shown to lead to opinion polarization in connected networks of sufficiently high modularity \cite{Banisch2018}. In the present approach, the effect of social interaction is directed towards the \textit{willingness} of or incentive for individuals to publicly express their opinion. We investigate the structural conditions that promote or hinder opinion expression of different opinion groups. This is firstly done from a game-theoretic angle. To address questions of bounded rationality and equilibrium selection, we also develop a dynamical systems perspective, using reinforcement learning in the form of $Q$-learning \cite{kianercy2012dynamics}. This allows us to perform a a mean-field approximation for the expected reward of the two opinion groups, which reduces the system to two dimensions.

In the following, we will first describe the baseline social structure and the two central structural parameters of the model. In section \ref{sec:game}, we represent the model as a multi-group majority game on the agent network, and investigate its Nash equilibria with respect to the structural parameters. Section \ref{sec:qlearning} introduces $Q$-learning and a subsequent two-dimensional approximation of the dynamical system. In section \ref{sec:bif} we perform a bifurcation analysis for the different parameters involved. We conclude with a discussion of the results and an outlook in section \ref{sec:discussion}.
\section{Social-structural setting\label{sec:2}}
For simplicity, we assume that there are two groups of individuals holding two different opinions on an issue. The opinion of an agent $i$, $o_{i}$, is referred to by either $1$ or $2$, depending on the group she belongs to. $G_{1}$ is the group of agents holding opinion 1, $G_{2}$ the one holding opinion 2. According to their opinion, the connections between agents are described by weighted blocks (the entries $q_{11}$, $q_{12}$, $q_{21}$ and $q_{22}$ in the different blocks represent the weight of every connection within that block)  according to the adjacency matrix $A$,
\begin{equation}
    A = \mymatrix.
    \label{eq:matrix}
\end{equation}
Opinion group $G_{1}$ has size $N_{1}$ and opinion group $G_{2}$ has size $N_{2}$. The weight of an edge between any two agents of community $1$ is $q_{11}$, and analogously $q_{22}$ for the second community. Cross-edge weights are given by $q_{12}$ and $q_{21}$. All weights $q_{11},q_{22},q_{12},q_{21} \in [0,1]$. We assume an undirected network, hence $$q_{12}=q_{21}.$$

The weights of the edges can be interpreted as the intensity or strength of the connections: The smaller the weight, the less strong agents notice the presence of each other. We can express the fraction of others an agent perceives to hold the same opinion by

\begin{equation}
    f_{11}=\frac{(N_{1}-1)q_{11}}{(N_{1}-1)q_{11}+N_{2}q_{12}}
    \label{eq:p11}
\end{equation}
for agents belonging to opinion group $G_{1}$ and
\begin{equation}
    f_{22}=\frac{(N_{2}-1)q_{22}}{(N_{2}-1)q_{22}+N_{1}q_{12}},
    \label{eq:p22}
\end{equation}
for agents that are part of opinion group $G_{2}$. The perceived fractions of others belonging to the other opinion group are consequently
\begin{equation}
    f_{12}=\frac{N_{2}q_{12}}{(N_{1}-1)q_{11}+N_{2}q_{12}} \label{eq:f12}
\end{equation}
and

\begin{equation}
    f_{21}=\frac{N_{1}q_{12}}{(N_{2}-1)q_{22}+N_{1}q_{12}}.\label{eq:f21}
\end{equation} We now introduce the two central structural parameters, $\gamma$ and $\delta$. They are the ratios of the weighted in-group to the out-group connections for each opinion group and given by\begin{equation}
\gamma=\frac{N_{1}-1}{N_{2}} \cdot \frac{q_{11}}{q_{12}} \label{eq:gam}\end{equation} and
\begin{equation}
\delta=\underbrace{\frac{N_{2}-1}{N_{1}}}_\text{group sizes}\:\:\:\cdot\:  \underbrace{ \:\:\frac{q_{22}}{q_{12}},}_\text{weights} \label{eq:del}
\end{equation} $\gamma>1$ or $\delta>1$ means that the agents of one opinion group are more strongly connected (under consideration of both the group sizes and the weights) to agents of the same than of the other opinion, while $\gamma<1$ or $\delta<1$ indicates that agents of the opinion group are more strongly connected to agents holding a different opinion. In the following, if we say that an opinion group is internally well-connected, we mean that the structural parameter of the group is bigger than 1.
With $\gamma$ and $\delta$, the above fractions (\ref{eq:p11}), (\ref{eq:p22}), (\ref{eq:f12}) and (\ref{eq:f21}) can be simplified to \begin{equation}
f_{11}=\frac{\gamma}{\gamma+1},\hspace{1cm}f_{12}=\frac{1}{\gamma+1},
\label{eq:2p11}
\end{equation}
\begin{equation}
f_{22}=\frac{\delta}{\delta+1},\hspace{1cm}f_{21}=\frac{1}{\delta+1}.
\label{eq:2p22}
\end{equation}
Alternatively, we can interpret the fractions $f_{11}$, $f_{22}$, $f_{12}$ and $f_{21}$ as the probabilities of interaction between agents of the different groups.\footnote{In this case, $A$ is interpreted as a stochastic block matrix and the weights $q_{11}$, $q_{22}$ and $q_{12}$ as probabilities of there being an edge between any two agents, depending on their group affiliation.} This interpretation will be made use of in section \ref{sec:qlearning}.

\section{A silence game \label{sec:game}}
We now use the social structure described in section \ref{sec:2} as the setting of a `silence game'. The opinions of the agents are fixed according to their group affiliation and do not change. Each agent can choose one of two actions: Public expression of opinion, or silence. Their preference over the actions depends on the perception of their opinion environment. If it appears to them that they are part of a minority opinion, they become silent. If they think that they are part of the majority, they express their opinion.\footnote{Games with fixed, different group affiliations of agents are considered e.g. in \cite{neary2011multiple} or \cite{NEARY2012301}.} But only the expressive agents shape the subjective impression of the opinion landscape of each individual. Silent ones do not contribute (they are not counted in the inequalities below). After all, silence means that the individual's opinion is not public.

Moreover, we introduce as an additional model assumption that opinion expression does not come for free. It is costly to express one's opinion,\footnote{We may think of the effort of typing a reply to someone in social media, or the effort of joining a demonstration for or against some issue.} which is accounted for by a constant cost $c$. This constant might make more than a simple (perceived) opinion majority necessary for an agent to also have an incentive to express her opinion.

The ordinal preferences of an individual $i$ over the actions $e$ (for opinion expression) and $s$ (for silence) are given as follows: The payoff or utility for opinion expression, $U_{i}(e,a_{-i})$ (given the actions of all others $a_{-i}$), is bigger than the one for silence,
\begin{equation}
    U_{i}(e,a_{-i})>U_{i}(s,a_{-i}), \nonumber
\end{equation} if more agents in the perceived social environment of $i$ speak out who share $i$'s opinion. 
We can make this condition explicit by the inequality 
\begin{equation}
    \frac{\sum_{\substack{j \in G_{1} \\  j\neq i}} q_{11} a_{j}}{(N_{1}-1)q_{11}+N_{2}q_{12}} > \frac{\sum_{j \in G_{2}} q_{12} a_{j}}{(N_{1}-1)q_{11}+N_{2}q_{12}}+c\label{eq:in1}
\end{equation}
if $i$ is part of opinion group $G_{1}$ and 
\begin{equation}
    \frac{\sum_{\substack{j \in G_{2} \\  j\neq i}} q_{22} a_{j}}{(N_{2}-1)q_{22}+N_{1}q_{12}} > \frac{\sum_{j \in G_{1}} q_{12} a_{j}}{(N_{2}-1)q_{22}+N_{1}q_{12}}+c\label{eq:in2}
\end{equation}
for $i$ being part of $G_{2}$. Both the right- and the left-hand side are normalized with respect to the overall weighted connections of agent $i$.
Here, the actions $a_{j}$ are given by $a_{j}=1$ for expression and $a_{j}=0$ for silence. If the two sides (\ref{eq:in1}) or (\ref{eq:in2}) are equal, the individual is indifferent in her preference over the actions.

A strategy profile is called a Nash equilibrium (NE) if no individual $i$ can increase her expected reward by unilaterally deviating from the equilibrium. In our system, the equilibrium condition is met if there is a strategy profile for which each individual that expresses herself has (\ref{eq:in1}) or (\ref{eq:in2}) (depending on the opinion group of the agent) satisfied, and if for each individual that is silent, the corresponding inequality is not fulfilled. 

It is already visible in (\ref{eq:in1}) and (\ref{eq:in2}) that apart from the fact that an individual does account for her own expressed opinion in the inequality ($i\neq j$ in the sum on the left-hand side), the rest of the contributions in the inequalities are the same for all agents of one opinion group. It is also visible that if (\ref{eq:in1}) or (\ref{eq:in2}) is satisfied for an agent $i$ that expresses herself, it must be satisfied for all silent individuals of her group as well: They `see' one more agent expressing their opinion than $i$, since $i$ does not account for herself in her evaluation of her environment. Hence, we have an additional positive term on their left-hand side. On the other hand, if the inequality is not fulfilled for a silent agent of one group, it can also not be fulfilled for an expressive one. Therefore, in a pure-strategy equilibrium, all agents of one opinion group must choose the same action.

This simplifies the inequalities above. If all agents of an opinion group act the same, (\ref{eq:in1}) and (\ref{eq:in2}) can be expressed in terms of the structural parameters $\gamma$ and $\delta$. Four pure-strategy NEs might be possible, depending on $\gamma$ and $\delta$. Both groups can be silent, or only one of them, but not the other, or none:

\begin{itemize}
\item If both groups express their opinion (we call this state $(e,e)$; the first entry stands for the collective action of $G_{1}$, the second for the action of $G_{2}$), the following conditions must be satisfied to make this state a NE: \begin{equation}
    \frac{(N_{1}-1) q_{11}-N_{2}q_{12}}{(N_{1}-1)q_{11}+N_{2}q_{12}}-c=\frac{\gamma-1}{\gamma +1}-c > 0,  \label{eq:ee1}
\end{equation}\footnote{We use equation (\ref{eq:gam}) in the equivalence.}
\begin{equation}
    \frac{(N_{2}-1) q_{22}-N_{1}q_{12}}{(N_{2}-1)q_{22}+N_{1}q_{12}}-c=\frac{\delta -1}{\delta +1}-c > 0.\label{eq:ee2}
\end{equation}
\item $(e,s)$ is a NE if
\begin{equation}\frac{(N_{1}-1)q_{11}}{(N_{1}-1)q_{11}+N_{2}q_{12}}-c=\frac{\gamma}{\gamma +1}-c > 0, \label{eq:es1}
\end{equation}
\begin{equation}
-\frac{N_{1}q_{12}}{(N_{2}-1)q_{22}+N_{1}q_{12}}-c=-\frac{1}{\delta +1} -c <0.\label{eq:es2}
\end{equation}
\item $(s,e)$ is a NE if
\begin{equation}
-\frac{N_{2}q_{12}}{(N_{1}-1)q_{11}+N_{2}q_{12}}-c=-\frac{1}{\gamma +1} -c <0, \label{eq:se1}
\end{equation}
\begin{equation}
    \frac{(N_{2}-1)q_{22}}{(N_{2}-1)q_{22}+N_{1}q_{12}}-c=\frac{\delta}{\delta +1}-c > 0. \label{eq:se2}
\end{equation}
\item $(s,s)$ is a NE if
\begin{equation}
    -c<0.
\end{equation}
\end{itemize}

\begin{figure}
\begin{center}
\includegraphics[width=8.6cm]{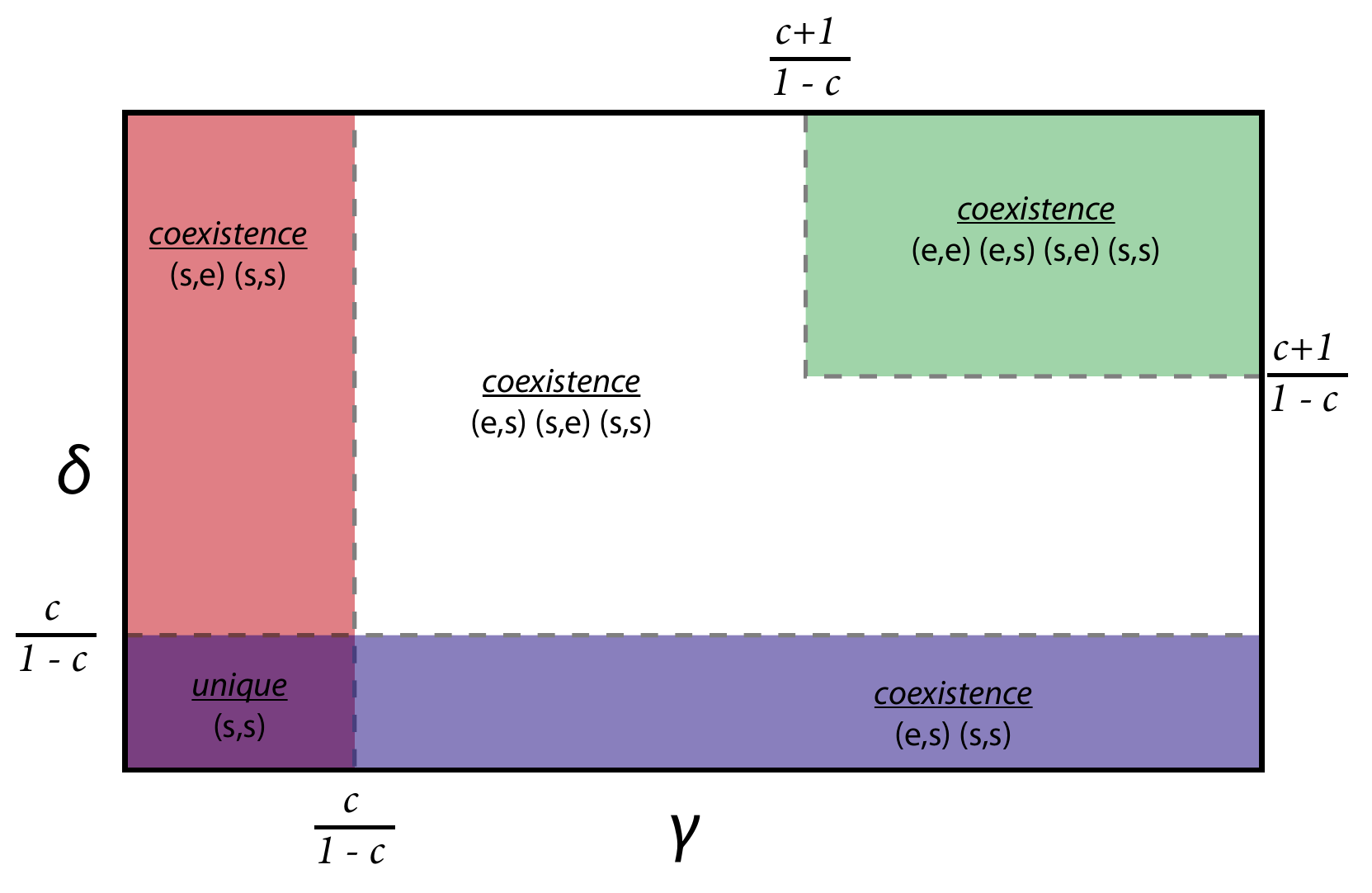}
\end{center}
\caption{The available pure-strategy Nash equilibria in different regimes of $\gamma$ and $\delta$. The equilibria are abbreviated by either $e$ for expression or $s$ for silence for each opinion group (the first entry is for the collective action of $G_{1}$, the second for the one of $G_{2}$). For costs $c>0$, $\gamma$ and $\delta$ below $\frac{c}{1-c}$ will lead to a situation in which the only available Nash equilibrium is one in which no one expresses her opinion publicly. An increase in the structural parameters above this threshold leads to additional Nash equilibria in which at least one of the two opinion groups speaks out. If both $\gamma$ and $\delta$ are bigger than $\frac{c+1}{1-c}$, an additional Nash equilibrium arises in which all agents express their opinion.}\label{fig:game}
\end{figure}

The different existence regimes of the pure-strategy NEs are given in Figure \ref{fig:game}. If $\gamma$ and $\delta$ are both smaller than $\frac{c}{1-c}$, then even if all group members express their opinion and the other opinion group is silent, it is too costly (compared to the amount of connections to agents of the own opinion group) to express one's opinion and the only NE  is the one in which all individuals are silent.
If $\gamma$ or $\delta$ or both are bigger than $\frac{c}{1-c}$, but smaller than $\frac{c+1}{1-c}$, either both opinion groups are silent or one of the groups expresses themselves, but not both: The strength of internal connections of each group are not sufficient to account for the negative influence of the other, expressive group. Not both (\ref{eq:ee1}) and (\ref{eq:ee2}) can be satisfied. Hence, this structural regime only allows public opinion predominance of one group (or complete silence).\footnote{If either only the conditions for $(e,s)$ or only for $(s,e)$ are satisfied, it is clear which opinion will dominate publicly (if any). If both are satisfied, the situation becomes more interesting in the sense that it depends on the initial conditions and the dynamical development of the system which opinion will predominate. We will approach these issues in sections \ref{sec:qlearning} and \ref{sec:bif}.}
If $\gamma$ and $\delta$ are both bigger than $\frac{c+1}{1-c}$, it is possible that both opinion groups express their opinon publicly at the same time. Then, the positive influence of the in-group members still dominates, even if all out-group members are expressive as well. Hence, also (\ref{eq:ee1}) and (\ref{eq:ee2}) are satisfied.

Obviously, there are also mixed-strategy NEs. Suppose the situation is as follows: The agents of each group mix their actions uniformly such that each agent is exactly indifferent between expressing herself or staying silent. Then, no one has an incentive for action change, and we therefore have a NE. This equilibrium is, nevertheless, only metastable in the sense that it only takes one agent to increase (or decrease) her expression probability in order to make it favourable for all other agents of one opinion group to express themselves (or become silent).

\begin{figure}
    \centering
        \includegraphics[width=8.6cm]{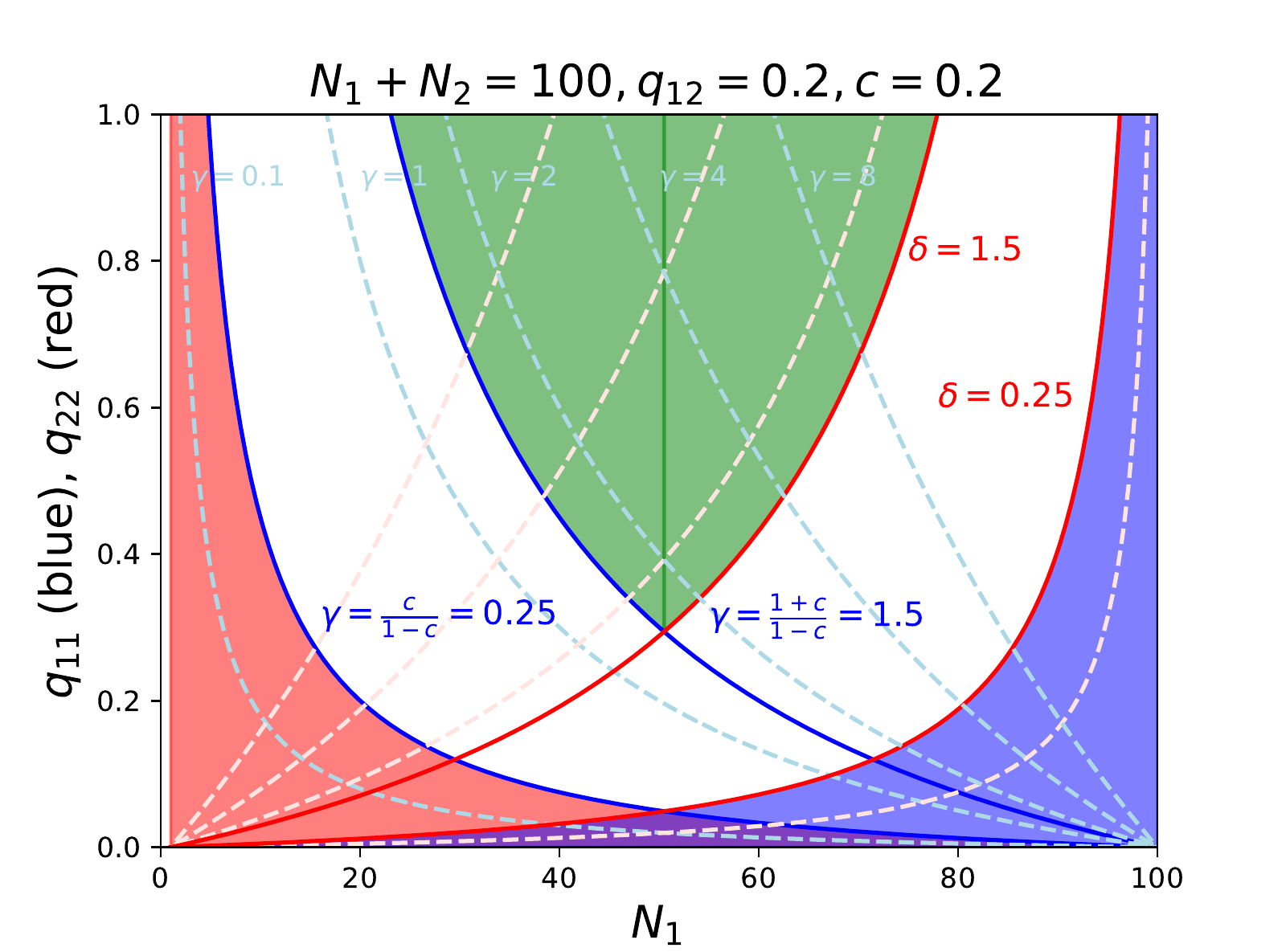}
    \caption{The constant $\gamma$- and $\delta$-curves for $N=100$ agents. They are plotted with respect to $N_{1}$, $N_{2}=N-N_{1}$, and $q_{11}=q_{22}$. Each blue curve stands for a combination of the number of opinion group members $N_{1}$ and internal connection weights $q_{11}$ that yields a constant value of the structural parameters $\gamma$, each red one for a combination of $N_{2}$ and $q_{22}$ that produces constant $\delta$. The color-coding for the different Nash equilibrium regions is analogous to Figure \ref{fig:game}. It is visible that the numerical minority of an opinion group cannot always be compensated by increasing $q_{11}$ (or $q_{22}$), the weight of a connection between two agents of the same opinion group. Moreover, the fixed $\gamma$- and $\delta$-curves are symmetric with respect to $N_{1}=N_{2}=50$, where they intersect. (For better readability, the dashed $\delta$-curves have not been labelled. They correspond to their $\gamma$-counterparts.)}
    \label{fig:my_label}
\end{figure}

$\gamma$ and $\delta$ do not only depend on the number of agents holding one or another opinion. They are also influenced by the internal connection weights of agents of one opinion group. Hence, a well-connected minority group can dominate public discourse if the corresponding structural parameter is above the threshold of $\frac{c}{1-c}$.
But while the regimes of different NEs in Figure \ref{fig:game} are displayed correctly, it might give the impression that $\gamma$ or $\delta$ are parameters that can be tuned by simply increasing the weight of a connection between two agents of the same group, that is, $q_{11}$ or $q_{22}$ (all other parameters fixed). That is not the case. Some numerical minorities cannot be balanced by increasing internal connections since $q_{11}$ and $q_{22}$ are bounded by $1$. If there are too few agents in one opinion group, even setting $q_{11}$ or $q_{22}$ to $1$ will not be elevate $\gamma$ or $\delta$ above a certain threshold. This is made visible in Figure \ref{fig:my_label}. The figure shows the different existence regimes of the NEs for different combinations of internal connection weights $q_{11}$ and $q_{22}$ and partitions of a total of $N=100$ agents between groups $G_{1}$ and $G_{2}$. $q_{12}$ and $c$ are fixed. Each point in the plot stands for a combination of the number of agents in opinion group $G_{1}$, $N_{1}$, and the in-group connection weights $q_{11}$, out of which one can compute the value of $\gamma$. The lines of constant $\gamma$ are plotted in red. Since the overall number of agents $N=100$ is fixed, $N_{2}$ is not independent and determined by the choice of $N_{1}$ by $N-N_{1}$. If we just assume that $q_{22}=q_{11}$, each point in the plot at the same time represents also a combination of the relevant parameters of opinion group $G_{2}$ out of which one can compute $\delta$. Curves of constant $\delta$ are the blue lines and symmetrical to the $\gamma$-curves with respect to $N_{1}=50$.

A vertical line in the plot, e.g. at $N_{1}=20$, can be interpreted as follows: Each constant $\gamma$ or $\delta$ value that it intersects on its way to $q_{11}=q_{22}=1$ is reachable for this partition of agents in the two groups if $q_{11}$ and $q_{22}$ are tuned accordingly. But if there is no intersection for a specific $\gamma$ or $\delta$, then even if the internal connection weights are maximized, the structural strength of the respective group cannot reach that value due to their limited group size.
For $N_{1}=20$, a state in which both opinion groups are expressing themselves (the upper right, green area in Figure \ref{fig:game}) cannot be reached since opinion group $G_{1}$ has too few agents to produce a $\gamma$ high enough to satisfy (\ref{eq:ee1}). In general, there are numerical thresholds (dependent on the costs $c$, the cross-group connection weight $q_{12}$ and the overall number of agents $N$) below which reaching a state in which both group express themselves or in which the own group becomes dominant becomes impossible from a game-theoretic perspective. The game-theoretic approach hence can give (all other parameters fixed) limits for the effect of group-internal coordination in the form of internal cohesion on public discourse.

\section{Q-learning and a dynamical systems perspective \label{sec:qlearning}}
While we are able to determine the Nash equilibria of the system, the game-theoretical point of view does not answer questions of equilibrium selection or the effects of bounded rationality. In this section, we will introduce a dynamical systems perspective to approach those questions. 

We posit a simple interaction mechanism between the agents on the network of section \ref{sec:2}. It is given as follows: If an agent expresses her opinion, she will be paired with a random neighbor. Now, the fractions $f_{11}$, $f_{12}$, $f_{21}$, and $f_{22}$ correspond to the \textit{probability} of meeting a neighbor of a certain opinion group given the own opinion group of an agent. The neighbor then gives (if she also is in an expressive state) social feedback to the agent, either agreement or disagreement, which will contribute to the agent's impression of her opinion environment. Put in an algorithmic way:
\begin{enumerate}
\item A random agent is selected.
\item If willing to speak out, the agent expresses her opinion to a random neighbor at cost $c$.
\item If the neighbor is also willing to speak out, she gives feedback on the agent's opinion.
\item According to the feedback, the agent will become more/less willing to speak out.
\end{enumerate}

As in \cite{Banisch2018}, we will describe the development of the system as reinforcement learning dynamics, more specifically, as dynamics induced by $Q$-learning. In $Q$-learning, the reinforcement mechanism that updates the agent's willingness to express her opinion is given by 
\begin{equation}
    Q_{i}^{t+1}=(1-\alpha)Q_{i}^{t} + \alpha r_{i}^{t},
    \label{eq:ql}
\end{equation}
where $r_{i}^{t}$ is the reward for agent $i$ at time step $t$ upon expression \begin{equation}
    r_{i}^{t}=\left\{
\begin{array}{lll}
-c &\text{for random neighbor being silent,} \\
-1-c& \, \text{for disagreeing random neighbor,} \\
1-c& \, \text{for agreeing random neighbor.}\\
\end{array}
\right.
\end{equation}
The $Q$-function is expected to converge to the reward over time.\footnote{(\ref{eq:ql}) describes Q learning for myopic agents, i.e. with discount factor 0.} The probability of expression is a function of the value of $Q_{i}$. We assume here a Boltzmann action selection mechanism, i.e. the probability of expression of agent $i$ is given by
\begin{equation}
    p_{i}^{t}=\frac{1}{1+e^{-\beta Q_{i}^{t}}}, \label{eq:boltzmann}
\end{equation}
the probability of staying silent by $1-p_{i}^{t}$. If $\beta=0$, the action choice of the agent is completely independent of the $Q$-values and randomized. For increasing $\beta$, the agent becomes more sensitive in her action selection towards her current evaluation of her local opinion environment. Then, a positive $Q$-value indicates that it is more likely for her to express herself than not, while a negative one indicates the opposite. If $\beta \rightarrow \infty$, the probabilities of the actions become deterministic. 

The expected reward for agent $i$ upon opinion expression is given by either (if $i$ belongs to opinion group $G_{1}$)
\begin{eqnarray}
 \e{}_{p}[r_{i}^{t}]=&&-c + f_{11} \frac{1}{N_{1}-1}\sum_{\substack {j \in G_{1} \\ j\neq i}} \frac{1}{1+e^{-\beta Q_{j}^{t}}} - \nonumber \\
 && f_{12} \frac{1}{N_{2}}\sum_{\substack {j \in G_{2}}} \frac{1}{1+e^{-\beta Q_{j}^{t}}},\label{eq:generalexp1}
\end{eqnarray}
or (if $i$ belongs to opinion group $G_{2}$)
\begin{eqnarray}
 \e{}_{p}[r_{i}^{t}]=&&-c + f_{22} \frac{1}{N_{2}-1}\sum_{\substack {j \in G_{2}\\ j\neq i}} \frac{1}{1+e^{-\beta Q_{j}^{t}}} - \nonumber \\
 && f_{21} \frac{1}{N_{1}}\sum_{\substack {j \in G_{1}}} \frac{1}{1+e^{-\beta Q_{j}^{t}}}.\label{eq:generalexp2}
\end{eqnarray}
We follow \cite{kianercy2012dynamics}, where Q-learning in two-player two-action games is investigated, and take the continuous-time limit of the Q-learning equation (\ref{eq:ql}). We divide time there into intervals $\delta t$. We replace $t+1$ with $t+\delta t$ and $\alpha$ with $\alpha' \delta t$. This yields
\begin{equation}
    Q_{i}(t+\delta t)-Q_{i}(t)=\alpha'\delta t(r_{i}(t)-Q_{i}(t)) \nonumber
\end{equation} and hence
\begin{equation}
    \dot{Q_{i}}=\alpha' (r_{i}(t)-Q_{i}(t)).
    \label{eq:conttime}
\end{equation}
Over time, the difference of the largest and the lowest $Q$-value of an opinion group decays at least exponentially in expectation (see the Appendix for the estimation): 
$$\frac{d}{dt} (Q_{i\in G_{1}}^{\text{max}}-Q_{i\in G_{1}}^{\text{min}}) \leq -\alpha' (Q_{i\in G_{1}}^{\text{max}}-Q_{i\in G_{1}}^{\text{min}}), $$
$$\frac{d}{dt} (Q_{i\in G_{2}}^{\text{max}}-Q_{i\in G_{2}}^{\text{min}}) \leq -\alpha' (Q_{i\in G_{2}}^{\text{max}}-Q_{i\in G_{2}}^{\text{min}}). $$

That is, the $Q$-values of the agents of one group are expected to converge over time. This allows us to employ a mean-field approximation for the expected reward of the two opinion groups: We introduce the average Q-values for each opinion group\footnote{Note the slight abuse of notation here: From now on, the index of $Q$ and $p$ will not indicate single individuals any more, but the average $Q$-value and the corresponding expression probability of the different opinion \textit{groups}.}
\begin{equation}
    Q_{1}(t)=\frac{1}{N_{1}} \sum_{i \in G_{1}} Q_{i}(t),\hspace{1cm}Q_{2}(t)=\frac{1}{N_{2}} \sum_{i \in G_{1}} Q_{i}(t). \label{eq:qaverage}
\end{equation}
This means that we do not distinguish any more between the agents of the respective opinion groups. We assign them the average of their group's $Q$-value.  This simplification will have an effect on the probability of opinion expression for the individuals. Instead of averaging over each groups probability of expression, we simply insert the averaged $Q$-values into the equation:
\begin{equation}
\frac{1}{N_{1}}\sum_{j \in G_{1}} \frac{1}{1+e^{-\beta Q_{j}(t)}} \longrightarrow \frac{1}{1+e^{-\beta Q_{1}(t)}}=p_{1}(t),
\label{eq:meanfield}
\end{equation}
\begin{equation}
\frac{1}{N_{2}}\sum_{\substack {j \in G_{2}}} \frac{1}{1+e^{-\beta Q_{j}(t)}} \longrightarrow \frac{1}{1+e^{-\beta Q_{2}(t)}}=p_{2}(t).
\label{eq:meanfield2}
\end{equation}

The expected reward for the different opinion groups are given by the equations\footnote{$f_{11}$, $f_{12}$, $f_{21}$, and $f_{22}$ have been replaced according to equations (\ref{eq:2p11}) and (\ref{eq:2p22}) with $\frac{\gamma}{\gamma+1}$, $\frac{1}{\gamma+1}$, $\frac{\delta}{\delta+1}$, and $\frac{1}{\delta+1}$.}
\begin{equation}
    \e{}_{p} [r_{1}(t)]=-c + \frac{\gamma}{\gamma+1} p_{1}(t) - \frac{1}{\gamma+1} p_{2}(t),
\end{equation}
\begin{equation}
    \e{}_{p} [r_{2}(t)]=-c + \frac{\delta}{\delta+1} p_{2}(t) - \frac{1}{\delta+1} p_{1}(t),
\end{equation}
where the probabilities of expression for each group are $p_{1}(t)$ and $p_{2}(t)$, and it is not distinguished any more between the individuals.

We can therefore write our two-dimensional formulation as follows:
\begin{equation} 
\dot{Q}_{1}(t) = \alpha' ( -c + \frac{\gamma}{\gamma+1} p_{1}(t) - \frac{1}{\gamma+1}p_{2}(t)-Q_{1}(t)),
\label{eq:Q1dot}
\end{equation}
\begin{equation} 
\dot{Q}_{2}(t) = \alpha' ( -c + \frac{\delta}{\delta+1} p_{2}(t) - \frac{1}{\delta+1}p_{1}(t)-Q_{2}(t)).
\label{eq:Q2dot}
\end{equation}

\begin{figure}
\includegraphics[width=8.6cm]{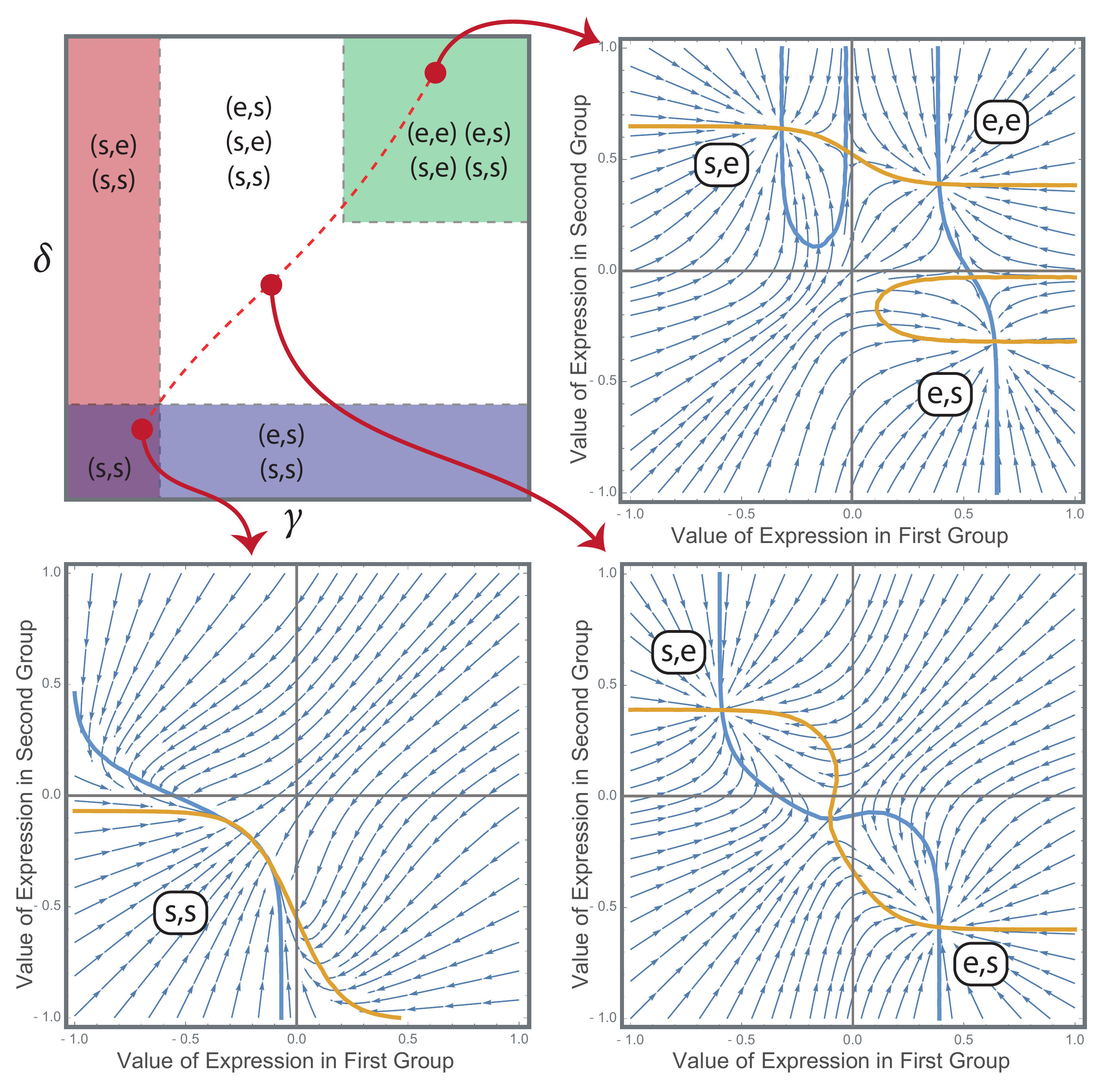}
\caption{Three phase portraits of the $Q_{1}$- (x-axis) and $Q_{2}$-values (y-axis) of the two-dimensional system for different configurations of $\gamma$ and $\delta$. We have $c = 0.1$, $\beta = 10$, and structural parameters $\gamma =\delta = 0.1$ (bottom left),  $\gamma =\delta = 1$ (bottom right), and $\gamma =\delta = 3$ (top right). The yellow and blue lines in the phase portraits are the isoclines of the equations for $Q_{1}$ and $Q_{2}$. The fixed points are located at their intersections.}
\label{fig:PhasePortrait}
\end{figure}

According to equations (\ref{eq:Q1dot}) and (\ref{eq:Q2dot}), we can produce a phase portrait of the system including its trajectories and fixed points for given exploration rate $\beta$, structural parameters $\gamma$ and $\delta$, and costs of expression $c$. An example of how the phase portraits change with $\gamma$ and $\delta$ is given in Figure \ref{fig:PhasePortrait}. 

\begin{figure}
    \centering
    \includegraphics[width=7cm]{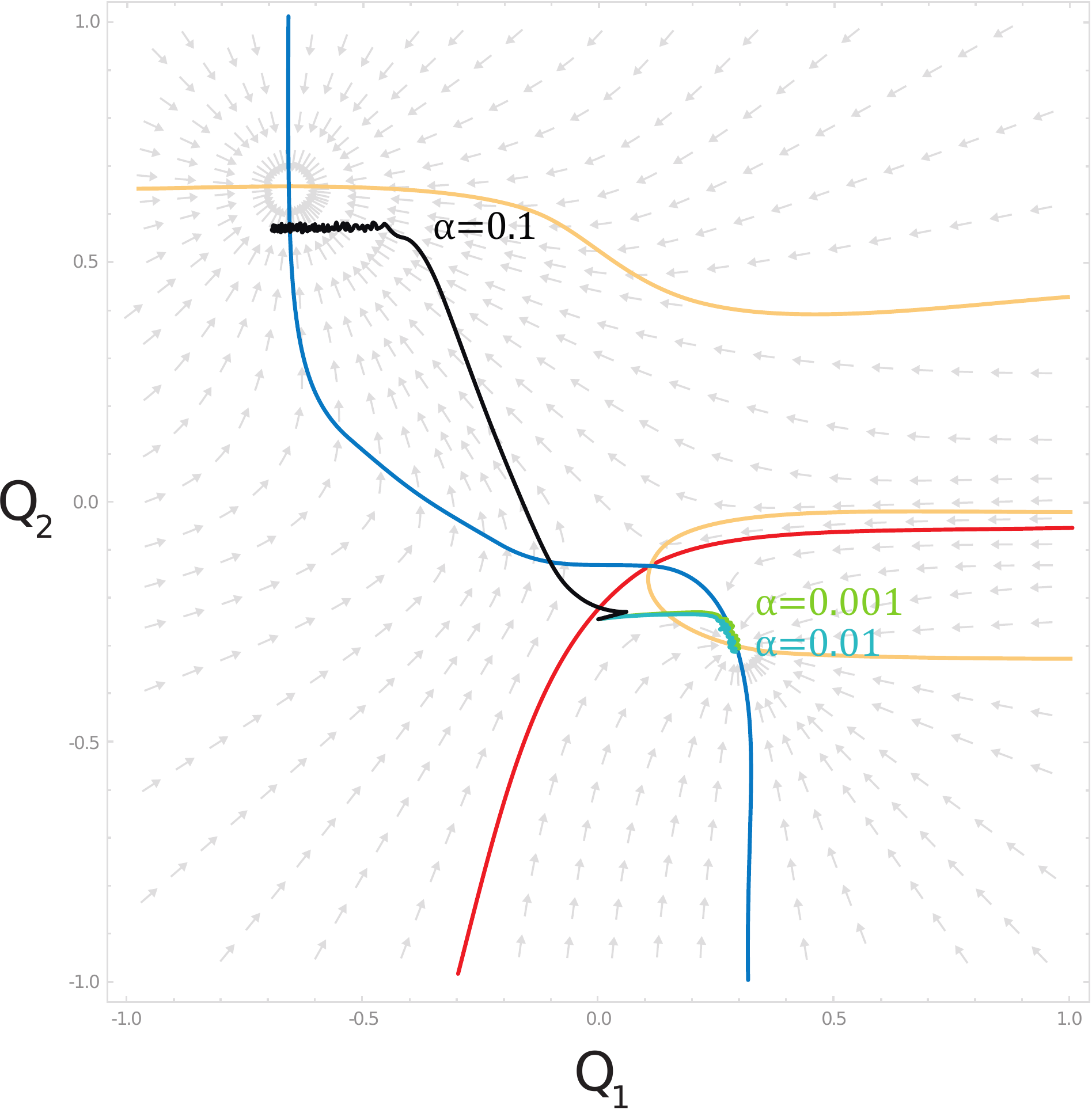}
    \caption{The trajectories of the $Q$-values in simulations, averaged over 50 runs with $N\cdot 10^5$ steps, for different values of $\alpha$ with a starting point close to the border (red line) of the two basins of attraction of the two stable fixed points. The starting $Q$-values were $Q_{i\in G_{1}}=0$, $Q_{i\in G_{2}}=-0.25$. There were $N=200$ agents, 100 of each opinion group, and $c=0.1$, $q_{11}=0.04$, $q_{12}=0.05$, and $q_{22}=0.15$. A relatively big $\alpha=0.1$ makes the trajectory leave the lower right basin of attraction of the two-dimensional system (black trajectory). Due to the high $\alpha$, the fixed point of the other basin of attraction is also missed by some margin. The lower $\alpha$, the closer the trajectories get to the fixed point and the more probable it is that they will stay in the basin predicted by the two-dimensional approximation. For $\alpha=0.01$ (turquoise) and $\alpha=0.001$ (light green), the trajectories run towards the predicted fixed point. The yellow and blue lines are the isoclines of the equations for $Q_{1}$ and $Q_{2}$. The fixed points are located at their intersections.}
    \label{fig:trajectories}
\end{figure}

There, it is visible that the stable fixed points of the system include basins of attraction, that is, regimes of values of $Q_{1}$ and $Q_{2}$ for which the system is expected to end up in those fixed points. The basins of attraction in the two-dimensional approximation correspond exactly to those of the stochastic $N$-agent system in the limit $\alpha \rightarrow 0$. For larger $\alpha$, both fixed points and basins of attraction do not necessarily correspond to the two-dimensional approximation. We show averages over simulation runs for different values of $\alpha$ in Figure \ref{fig:trajectories}. 

\section{Bifurcation and stability analysis \label{sec:bif}}
In order to find the fixed points of $Q_{1}$ and $Q_{2}$, we set (\ref{eq:Q1dot}) and (\ref{eq:Q2dot}) to 0, solve (\ref{eq:Q1dot}) for $Q_{2}$ and insert it into (\ref{eq:Q2dot}), which yields:
\begin{equation}
    Q_{2}=-\frac{1}{\beta} \ln{(\frac{1}{\frac{\gamma}{1+e^{-\beta Q_{1}}}-(\gamma+1)(Q_{1}+c)}-1)}
    \label{eq:Q2}
\end{equation}
\begin{eqnarray}
 \frac{\delta}{\delta+1}&&(\frac{\gamma}{1+e^{-\beta Q_{1}}}-(\gamma +1)(Q_{1}+c))-  \frac{1}{\delta+1} \frac{1}{1+e^{-\beta Q_{1}}} + \nonumber \\ && \frac{1}{\beta}\ln{(\frac{1}{\frac{\gamma}{1+e^{-\beta Q_{1}}}-(\gamma+1)(Q_{1}+c)}-1)}-c=0
\label{eq:Q1}
\end{eqnarray}

Equation (\ref{eq:Q1}) now gives us the $Q_{1}$-value of the fixed points of the system, with which we can calculate the corresponding $Q_{2}$-value by equation (\ref{eq:Q2}). In essence, the fixed points depend on four parameters: $\beta$, $\gamma$, $\delta$, and $c$. We will carry out a bifurcation analysis of these parameters in the following.

After having solved equations (\ref{eq:Q1}) and (\ref{eq:Q2}) for $Q_{1}$ and $Q_{2}$, we can assess the stability of the respective fixed points by calculating the eigenvalues of their Jacobian; two negative (real parts of the) eigenvalues indicate a stable attractor. 
In the following, we analyze the bifurcation structure of the system depending on the different types of parameters in the system.

\subsection{Structural power} 
The parameter $\gamma$ describes the ratio of internal versus external connectedness of $G_{1}$. $\gamma >1$ means that on average each member of $G_{1}$ is connected to more agents of the own than of the other opinion group. (Everything stated in this paragraph applies equivalently to $\delta$, which is just the parameter for the ratio of internal versus external connectedness of the other group.)

\begin{figure}
\begin{center}
\includegraphics[width=8.6cm]{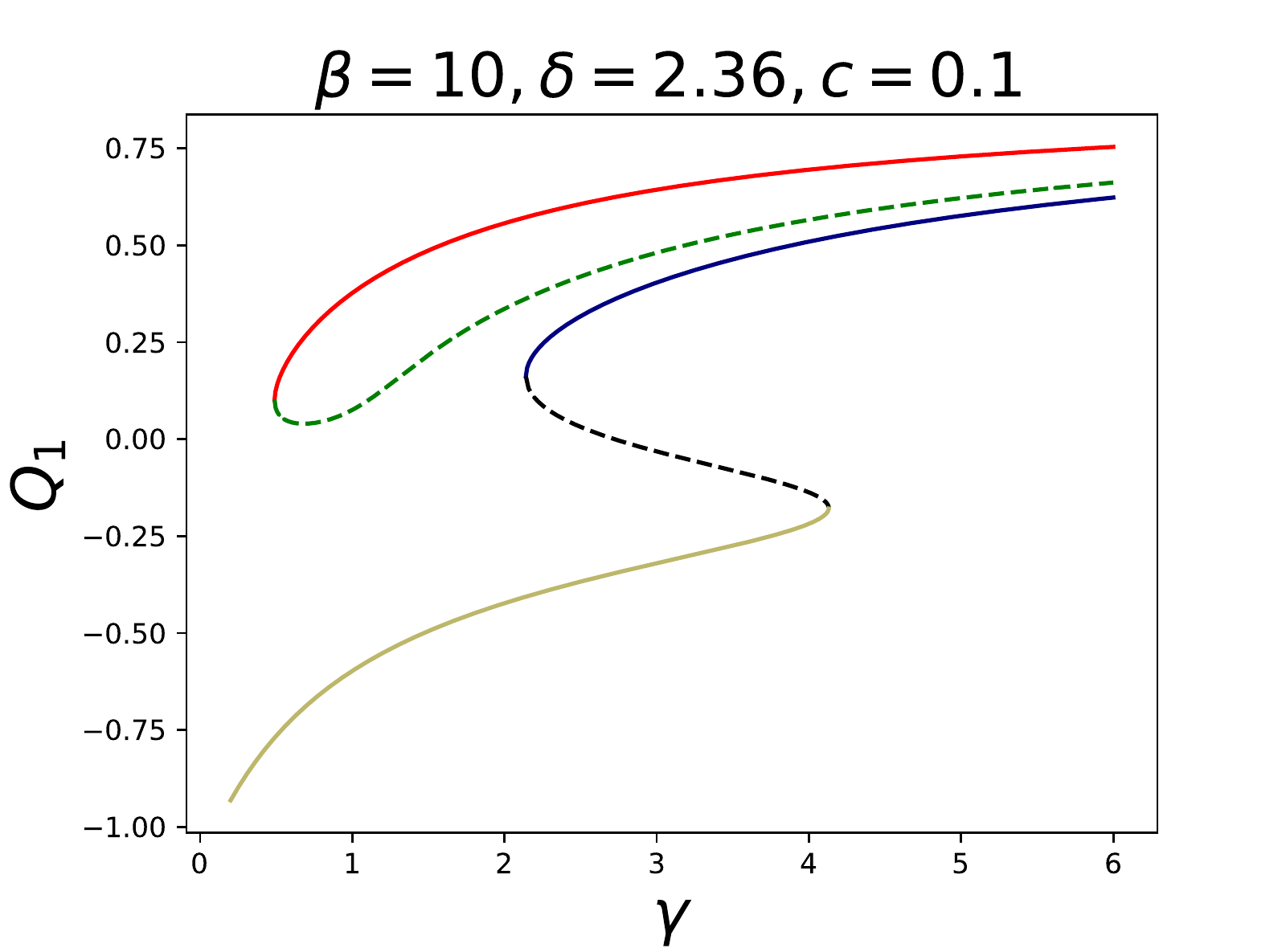}
\includegraphics[width=8.6cm]{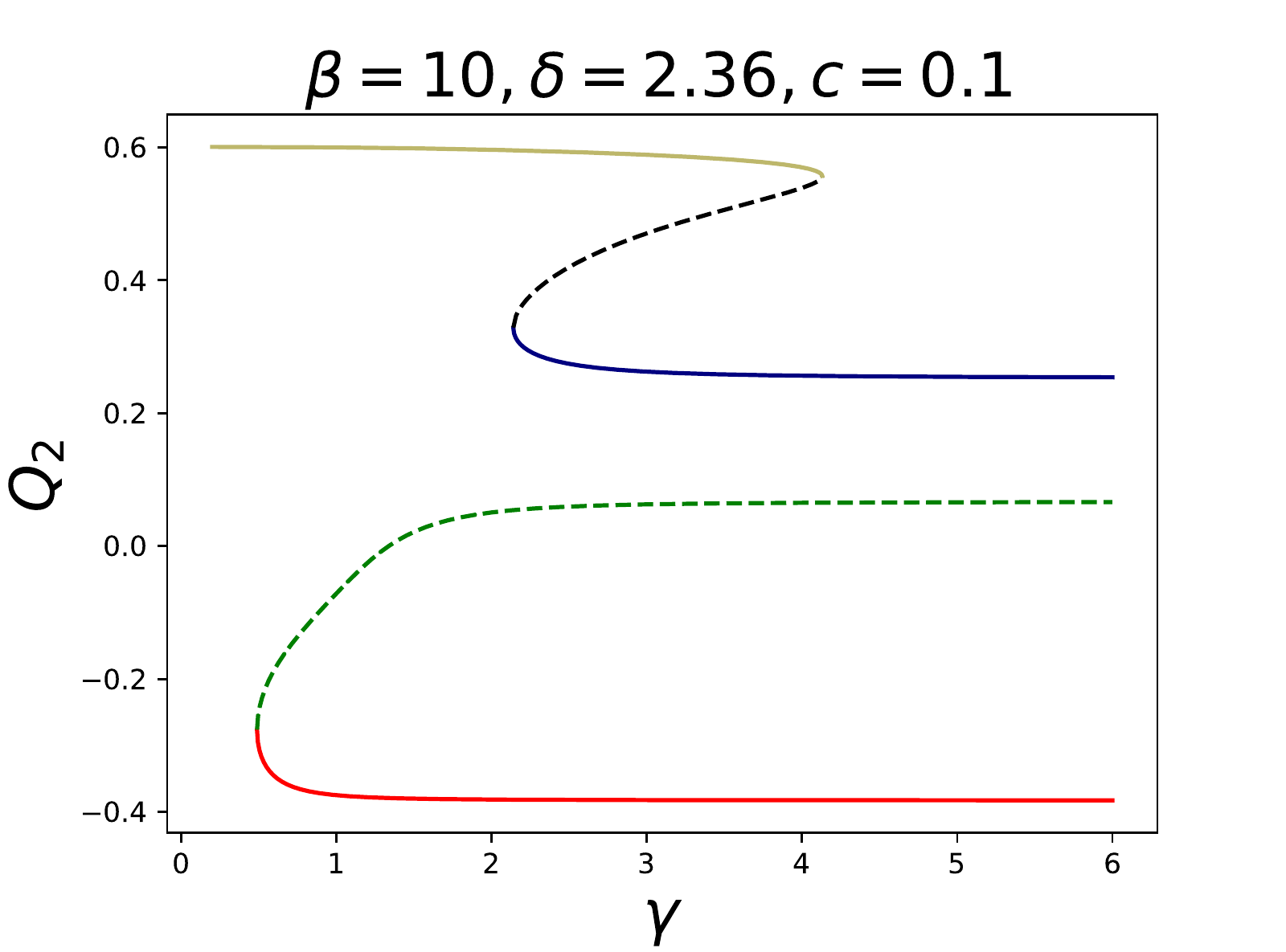}
\caption{The development of the $Q_{1}$- (top) and $Q_{2}$-value (bottom) of the fixed points with $\gamma$ given $\beta$, relatively high $\delta$, and $c$. The colors of the curves in the two plots indicate the different fixed point pairs of $Q_{1}$ and $Q_{2}$. A dashed line indicates an unstable fixed point, a continuous one a stable fixed point. It is visible in the plots that a poorly connected opinion group $G_{1}$ ($\gamma<0.5$) will be driven into silence by the other group (beige curve). With increasing in-group connectivity, fixed points arise for which $G_{1}$ expresses their opinion in two saddle-node bifurcations (red for an an $(e,s)$-equilibrium and blue for $(e,e)$). For $\gamma>4.5$, $G_{1}$ is so well-connected that the equilibrium disappears in which the group is silent.}
\label{fig:strpwhigh}
\end{center}
\end{figure}
As is visible in Figure \ref{fig:strpwhigh}, for small $\gamma$ ($<0.5$), given $\beta=10$, $\delta=2.36$ (that is, a quite well-connected opposite opinion group) and $c=0.1$, there is only one (stable) fixed point with negative $Q_{1}$-value and positive $Q_{2}$. While $\gamma$ grows, a saddle-node bifurcation occurs such that one stable and one unstable fixed point appear for positive $Q_{1}$ and negative $Q_{2}$. Another saddle-node bifurcation occurs at around $\gamma = 2$; and for $\gamma>4.2$, the low-$Q_{1}$ fixed points disappear in another saddle-node. 

How can this be interpreted? In essence, an opinion community that is not well-connected internally ($\gamma<0.5$) will be driven into silence (a Q-value much lower than 0) by the opposite opinion group that is internally more cohesive. With increasing $\gamma$, that is, increasing internal connectedness, other fixed points appear in which the former group is expressive. To be precise, the $Q$-values here are only indicative of probabilities of opinion expression according to the Boltzmann action selection which depends on $Q$. If $Q$ is smaller than 0, the probability of expression is smaller than the probability of staying silent. In the following, if we say that one opinion group is expressive, we mean that they have a $Q$-value bigger than 0 which makes their probability of expression higher than that of silence. With a further increase of $\gamma$, the stable fixed point for which only the opposite opinion group is expressive disappears and we remain with three fixed points (the middle one unstable), for which either the first opinion group is `loud' alone or both groups express their opinions. Hence, $G_{1}$ is now too cohesive to be driven into silence by the other group. Increased internal cohesion of one opinion group can hence have the effect that this group, which is not necessarily a majority, will dominate public discourse.
\begin{figure}
\begin{center}
\includegraphics[width=8.6cm]{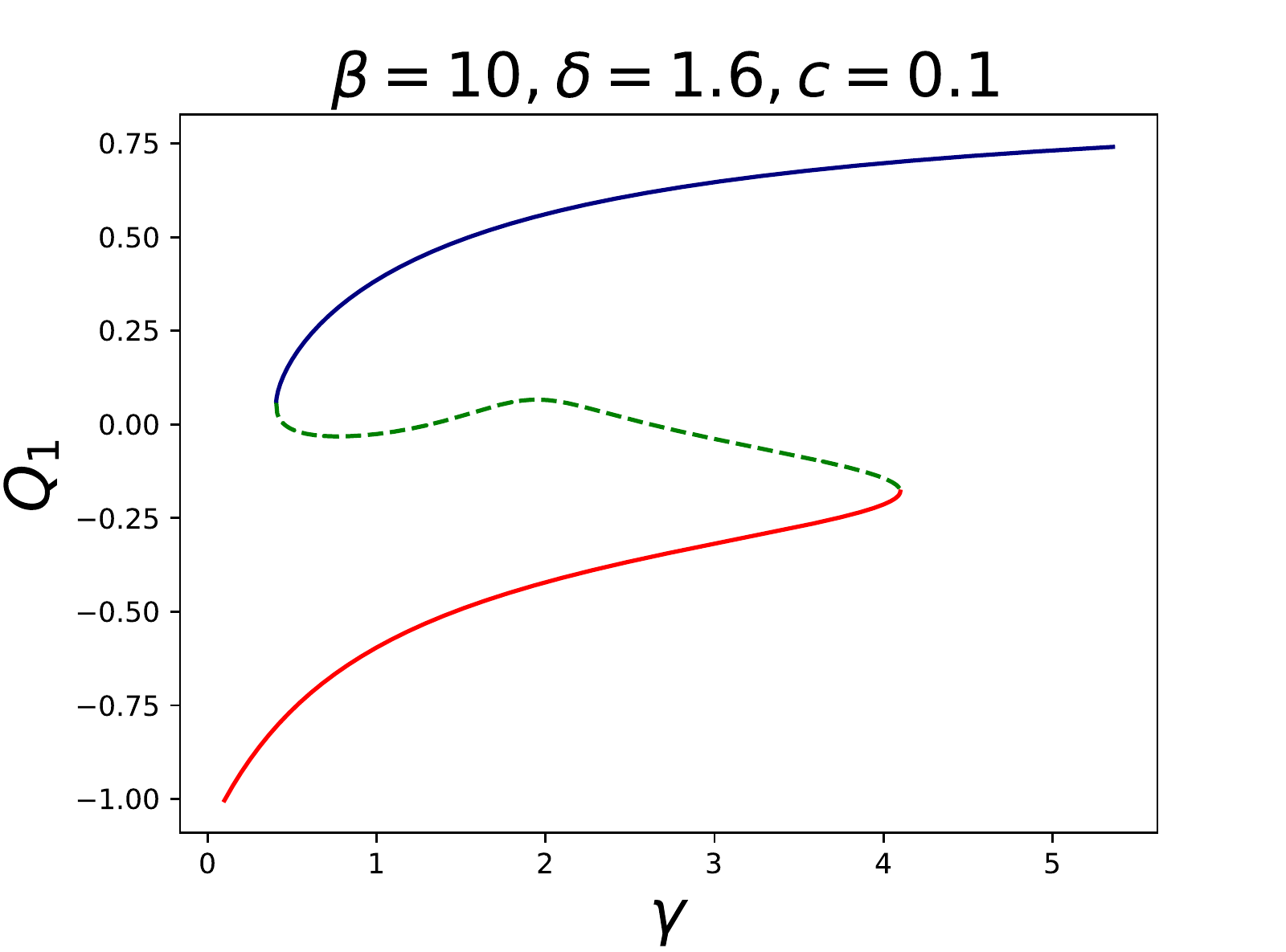}
\includegraphics[width=8.6cm]{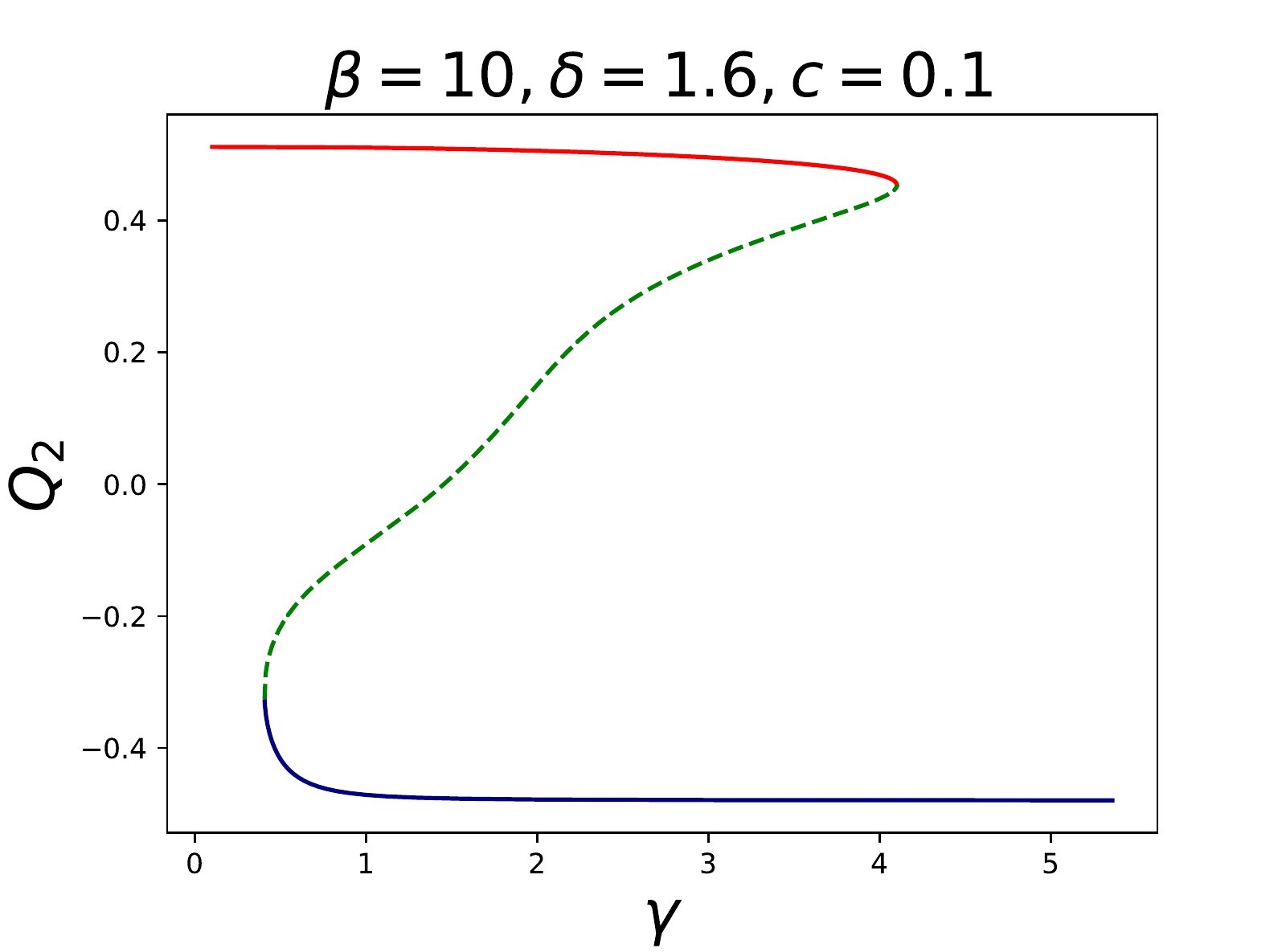}
\caption{The development of $Q_{1}$- and $Q_{2}$-fixed points with $\gamma$ given $\beta$, moderate $\delta$ and $c$. For $\gamma<0.4$, only group $G_{2}$ is expressive. A second fixed point arises for higher $\gamma$ in which $G_{1}$ is predominating public discourse. There is no fixed point in which both groups are expressive.}
\label{fig:strpwlow}
\end{center}
\end{figure}

A lower $\delta$-value (e.g. $\delta = 1.6$) leads to a reduction in available fixed points (Figure \ref{fig:strpwlow}) such that only two saddle-node bifurcations occur and at high $\gamma$ only one fixed point remains in which the first opinion group is expressive.

\subsection{Costs}

\begin{figure}
\begin{center}
\includegraphics[width=8.6cm]{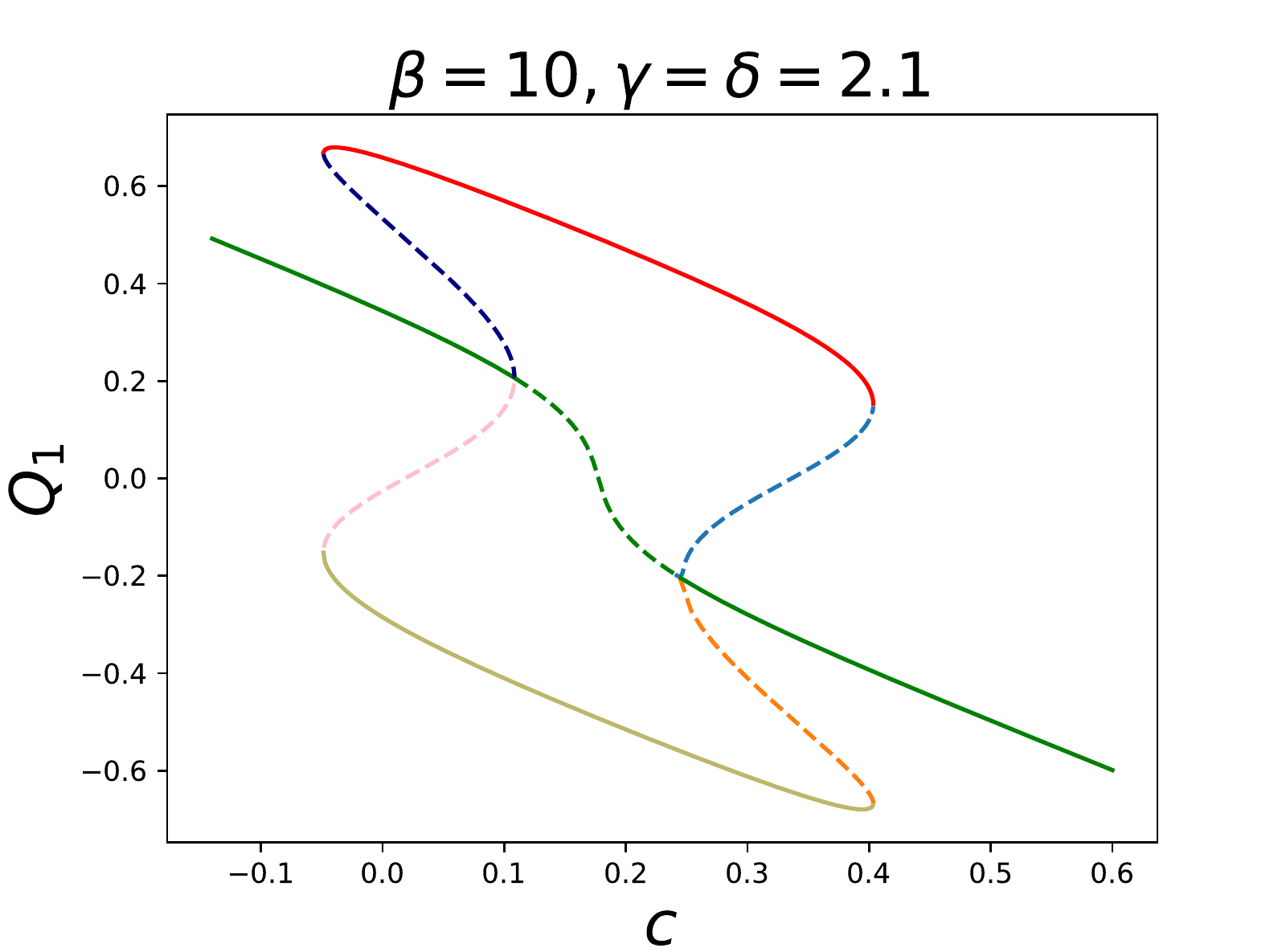}
\includegraphics[width=8.6cm]{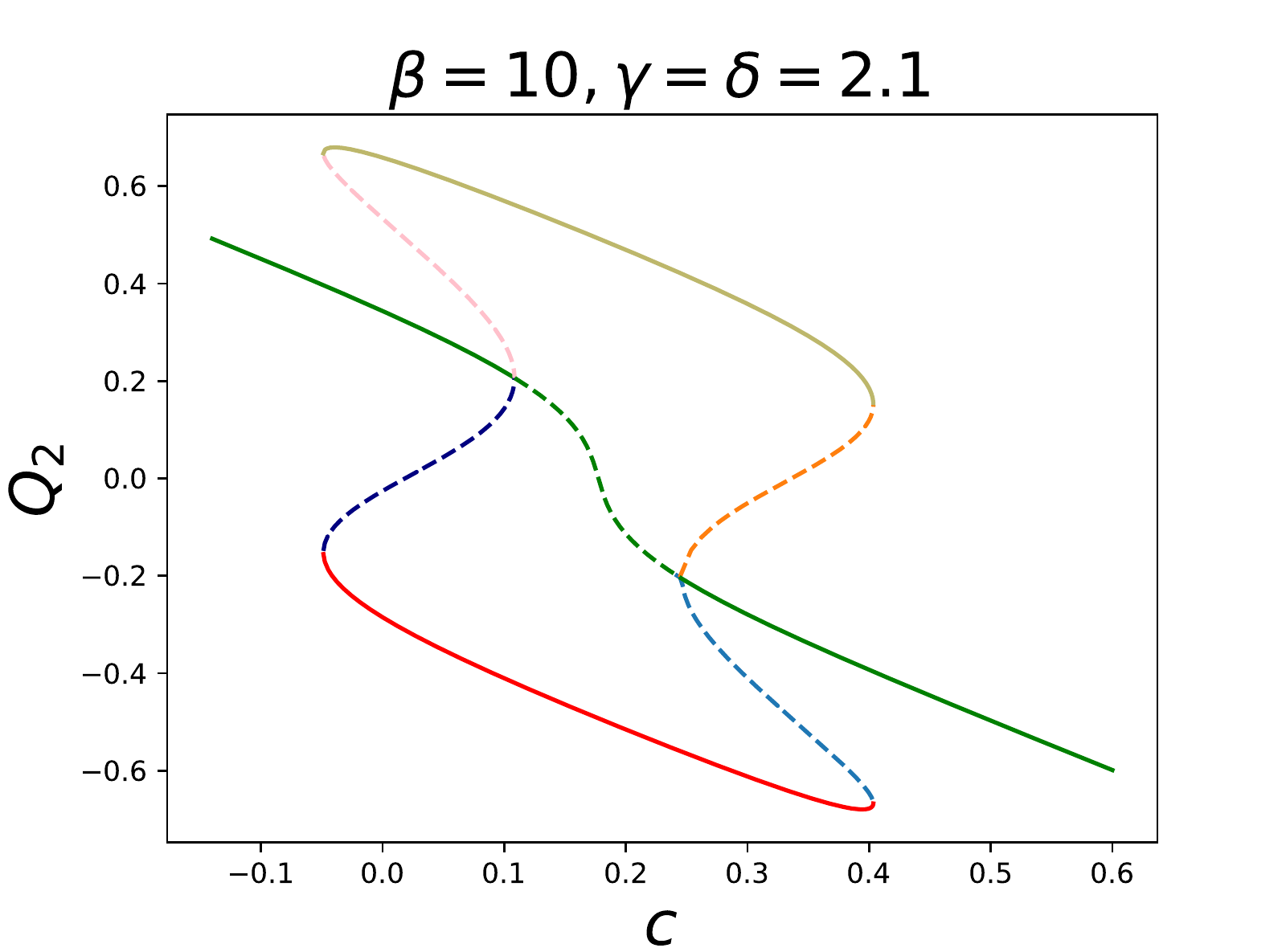}
\caption{The development of the fixed points with $c$ given $\beta$ and $\gamma=\delta$. The two are symmetric since $c$ is the same for both and has the same impact on both groups if they also have identical structural parameters. If expression is costly, everyone is silent, if it has negative costs, everyone speaks out.}\label{fig:costs}
\end{center}
\end{figure}

The costs for opinion expression have a profound impact on the fixed points of the system.
If opinion expression is very `expensive,' (in Figure \ref{fig:costs}: $c>0.4$), there is only one fixed point in the system for which both opinion groups stay silent. For decreasing costs, two pairs of fixed points arise in a saddle-node bifurcation. Each of the pairs corresponds to a situation in which one opinion group is expressive, while the other is silent (in Figure \ref{fig:costs}, we have identical values for $\gamma$ and $\delta$). 
The fixed point in which both opinion groups are silent becomes unstable with decreasing $c$ in a pitchfork bifurcation. Below $c=0.1$, another pitchfork bifurcation arises for which the stable fixed point now corresponds to a state in which both groups are expressing their opinion.
Costs can also be negative: Then, the individuals might be intrinsically motivated or externally encouraged to speak out.\footnote{Ideas such as e.g. free speech might have such an effect: People then see it as their duty to voice their opinion, \textit{especially} if it does not conform to the apparent majority.} For sufficiently negative costs (in the case of Figure \ref{fig:costs}: $c<-0.05$), only one fixed point exists: Everyone has an incentive to speak out, at least for internally well-connected opinion groups. The fixed points for which only one of the groups is expressive disappear in two saddle-nodes.

\subsection{Asymmetric costs}
The model allows us to also assign different costs to each opinion group, such that $c_{1}\neq c_{2}$.
Internal motivation for a cause, for example, can be an incentive to speak out and might even be indicated by negative costs (that is, an urge to express one's opinion). Moreover, there might be biases in the infrastructures on which debate takes place such that it takes more effort for one group to speak out than for the other.\footnote{One may think here about online platforms whose design favours engagement of certain demographic groups or states that encourage or try to prevent certain opinion groups to speak out.}

\begin{figure}
    \centering
    \includegraphics[width=8.6cm]{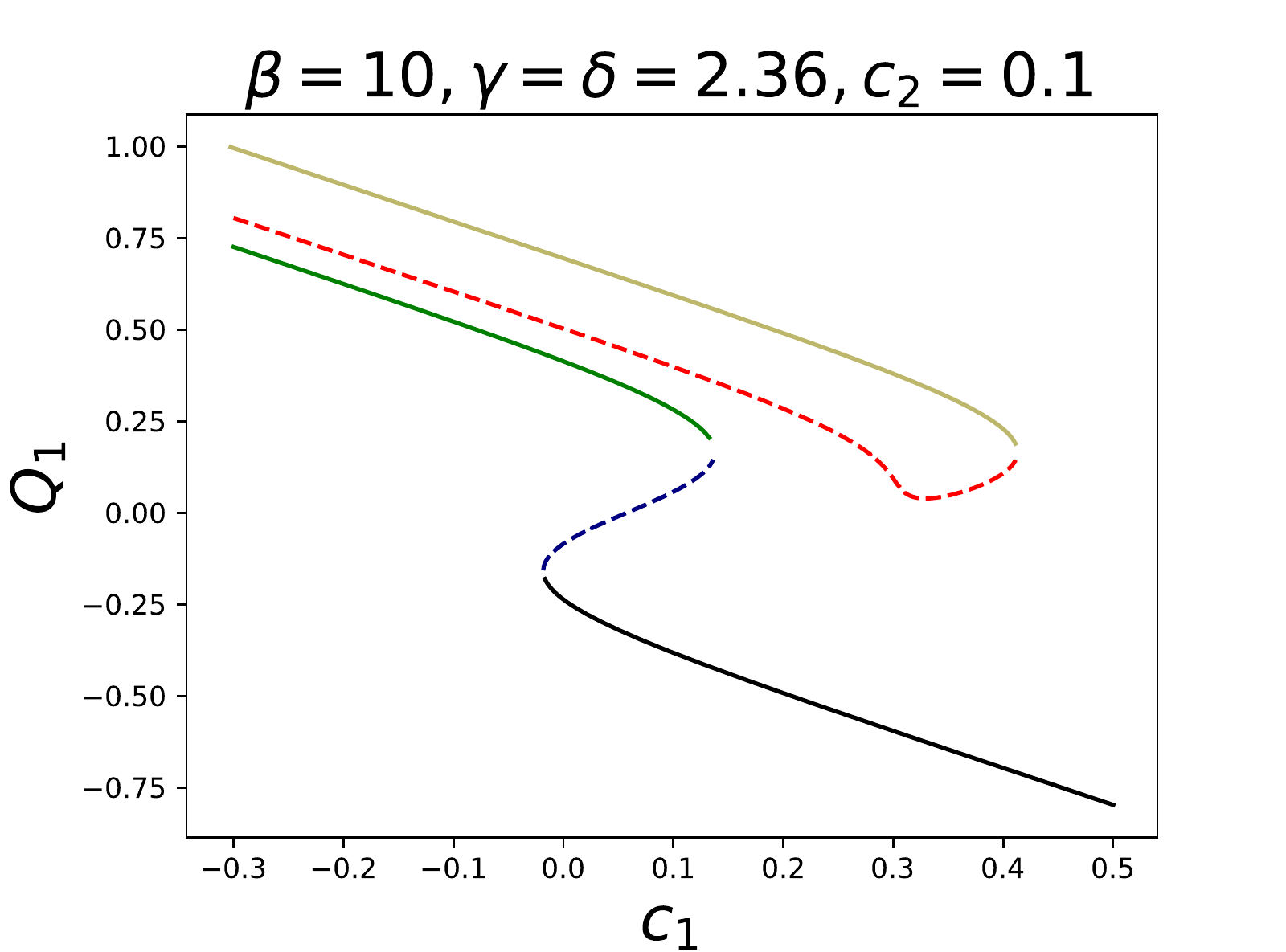}
    \includegraphics[width=8.6cm]{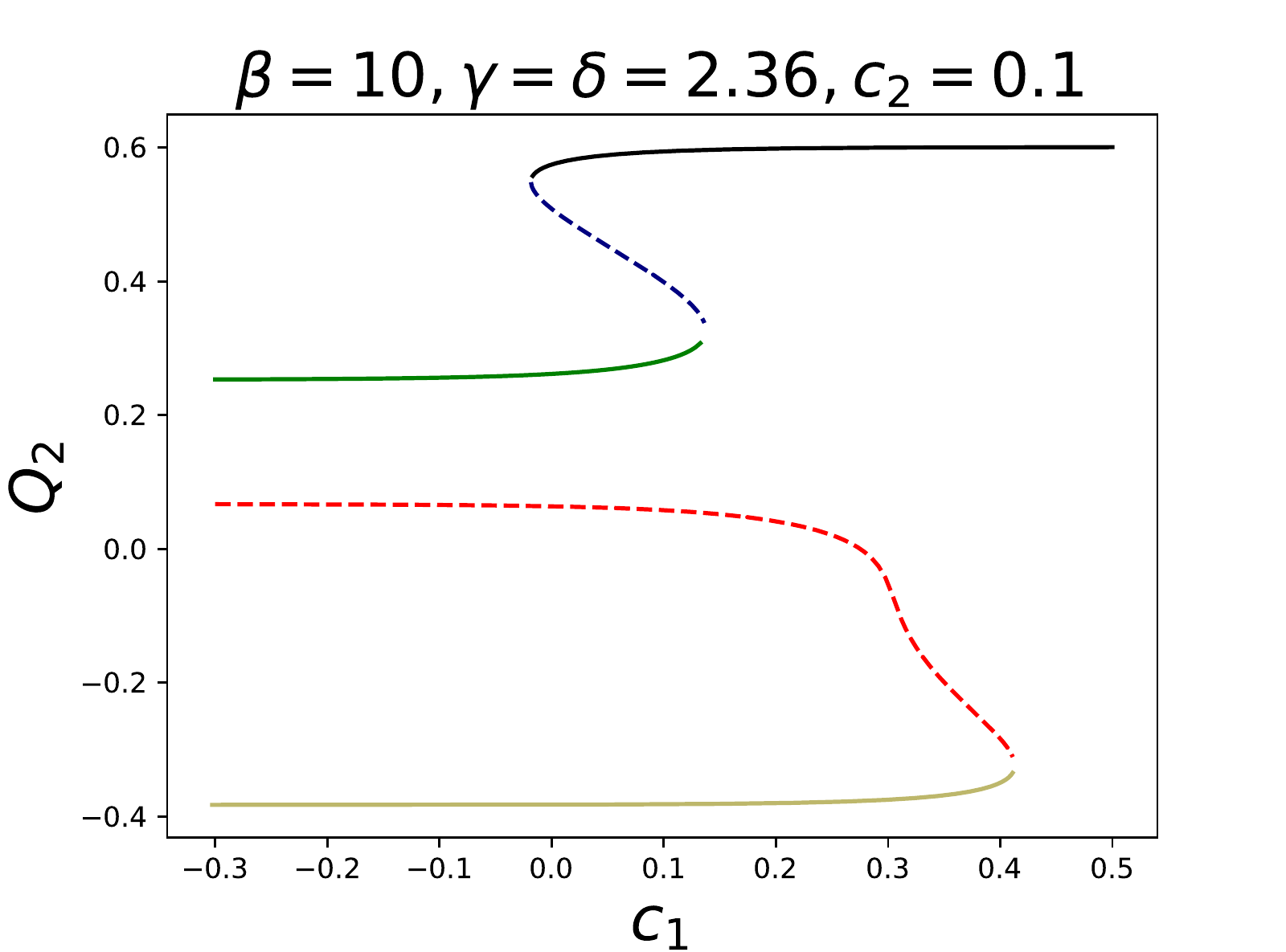}
    \caption{Fixed-point development with $c_{1}$ independent of $c_{2}$, given $\beta=10$, $\gamma=\delta=2.36$, $c_{2}=0.1$. Strongly negative $c_{1}$ corresponds to a strong motivational disposition (or many hard cores) in the opinion group to express their opinion. There, only fixed points in which this opinion group is expressive exist. For decreasing motivation (fewer hard cores), fixed points arise in which the second opinion group is the only expressive one.}
    \label{fig:asymcosts}
\end{figure}

The bifurcation in Figure \ref{fig:asymcosts} (for the case of two internally well-connected opinion groups) illustrates the effect that different expression costs in the populations exhibit on public discourse.
In Figure \ref{fig:asymcosts}, a bifurcation over $c_{1}$ is shown. Negative costs for opinion expression in opinion group $G_{1}$ yield two stable equilibria in which opinion group 1 is expressive, either together with opinion group $G_{2}$ or alone. With increasing costs, a stable fixed point arises in a saddle node for which $G_{1}$ is silent (at $c_{1} \approx 0$), while $G_{2}$ is expressive. 
At $c_{1} \approx 0.15$ and at $c_{1} \approx 0.4$, the two fixed points for which $G_{1}$ expresses opinion disappear. For costs that high, opinion group $G_{1}$ will not be publicly audible any more. Asymmetric costs can hence drive certain opinion groups into silence.

\subsection{Exploration rate}
The parameter $\beta$ determines how sensitive agents are in their actions towards the current evaluation of their expected reward. A high $\beta$-value indicates a choice of the agent similar to a best response to their current evaluation of the expected reward, while $\beta=0$ means that each available action is chosen with equal probability. 

\begin{figure}
\begin{center}
\includegraphics[width=8.6cm]{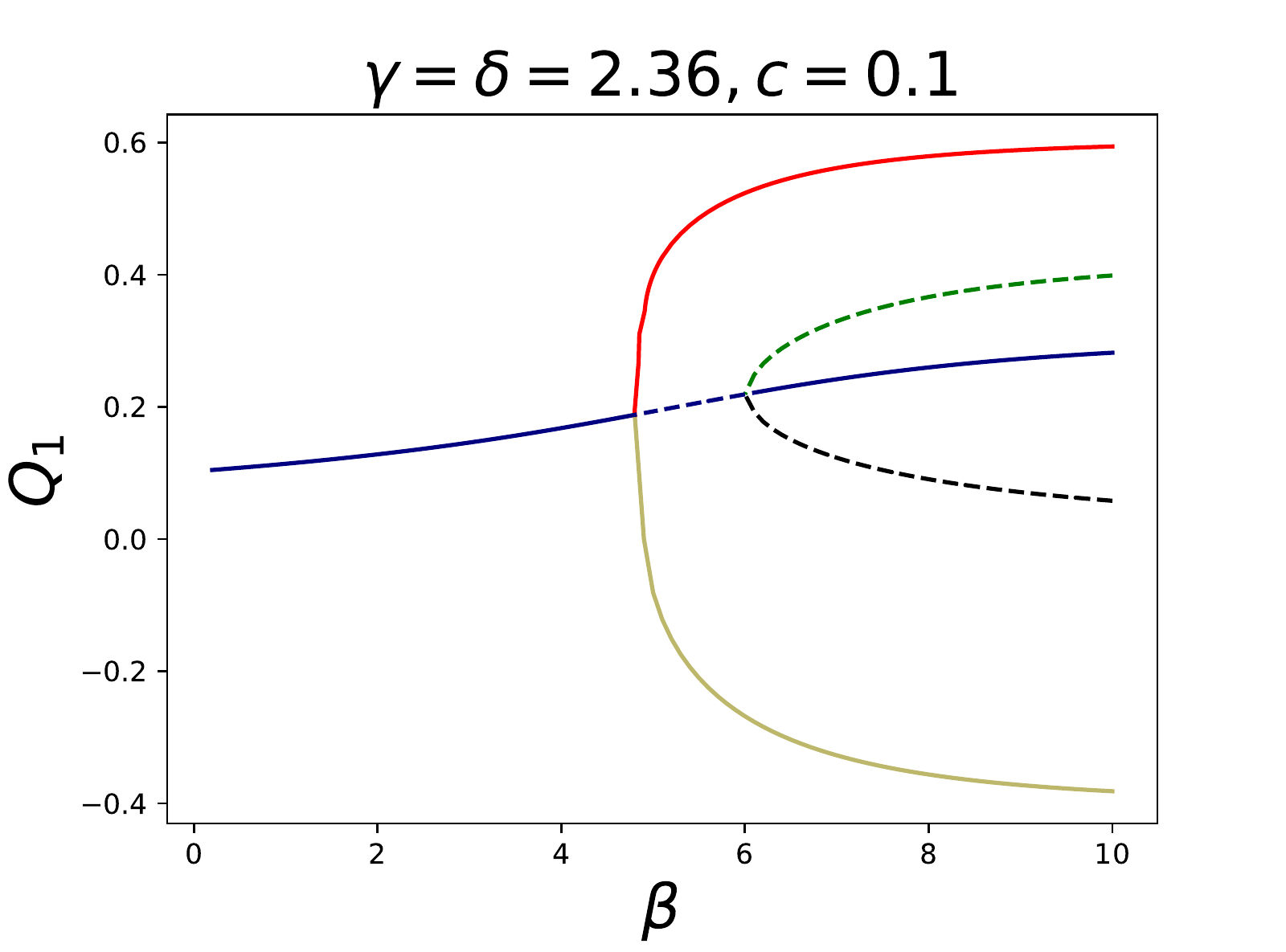}
\includegraphics[width=8.6cm]{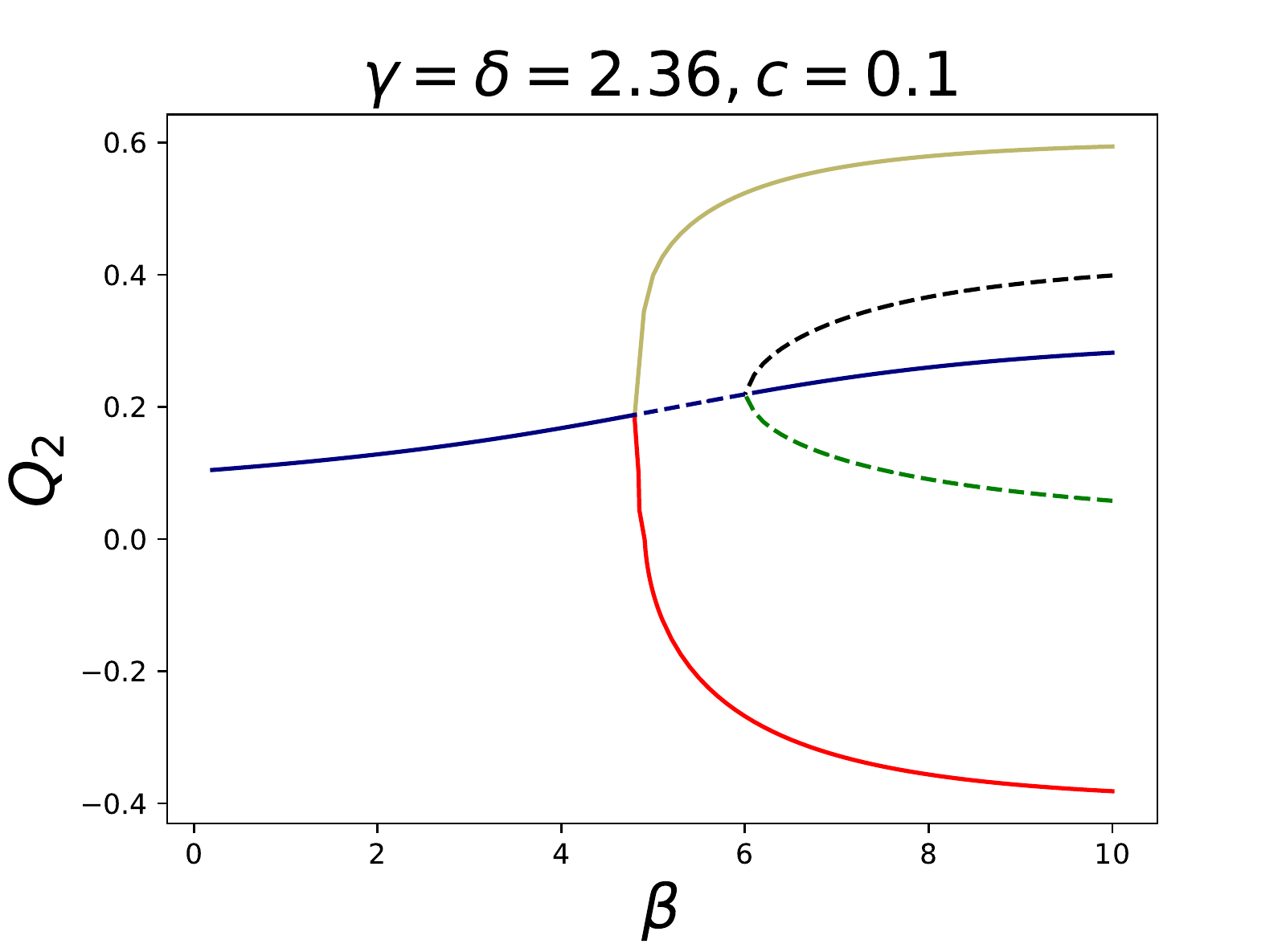}
\caption{The development of the fixed points with $\beta$ given $c=0.1$ and $\gamma=\delta=2.36$. Since $\gamma$ and $\delta$ are the same, the plots are symmetric.}\label{fig:beta}
\end{center}
\end{figure}

As is visible in Figure \ref{fig:beta}, for very low $\beta$, there is only one fixed point available with a very low $Q$-value for both opinion groups. With $\beta$ ($\approx 5$), further fixed points arise in a supercritical pitchfork bifurcation, and then, at $\beta >6$, another (now subcritical) pitchfork bifurcation arises, such that we arrive at three stable fixed points (one in which both groups are in an expressive mode, and one for opinion dominance for each group) and two unstable ones in-between. Hence, if the action selections is close to a best response, we get more possible equilibria in the system. In the intermediate region, we have a situation in which only one of the two groups can be expressive, despite them both being internally well-connected.

\section{Discussion and outlook \label{sec:discussion}}

\paragraph{The spiral of silence and beyond.} The present model provides a structural view on collective opinion expression. It reproduces the counterintuitive result postulated by Noelle-Neumann in her theory of the spiral of silence \cite{noelle2004spiral,NoelleNeumann1974}, namely the possibility of the public dominance of a minority opinion. While the influence of mass media has been stressed in many publications concerning the spiral of silence, we show that no mass media is needed for this effect. Being an internally well-connected community alone can be enough to gain public opinion predominance. This finding gains traction in light of the advent of social media, which facilitated communication among like-minded people and decentralized information distribution.
Apart from that, the present approach also provides conditions for the `overcoming' of the spiral of silence (in the sense that both groups express their opinion publicly), for which the numerical proportions do not necessarily have to change. The increase in internal cohesion of the different opinion groups can be sufficient. On the other hand, it is also shown that if the minority is too small, even maximum internal cohesion cannot heave the minority opinion into public predominance (see again Figure \ref{fig:my_label}). 

\begin{figure}
    \centering
    \includegraphics[width=8.6cm]{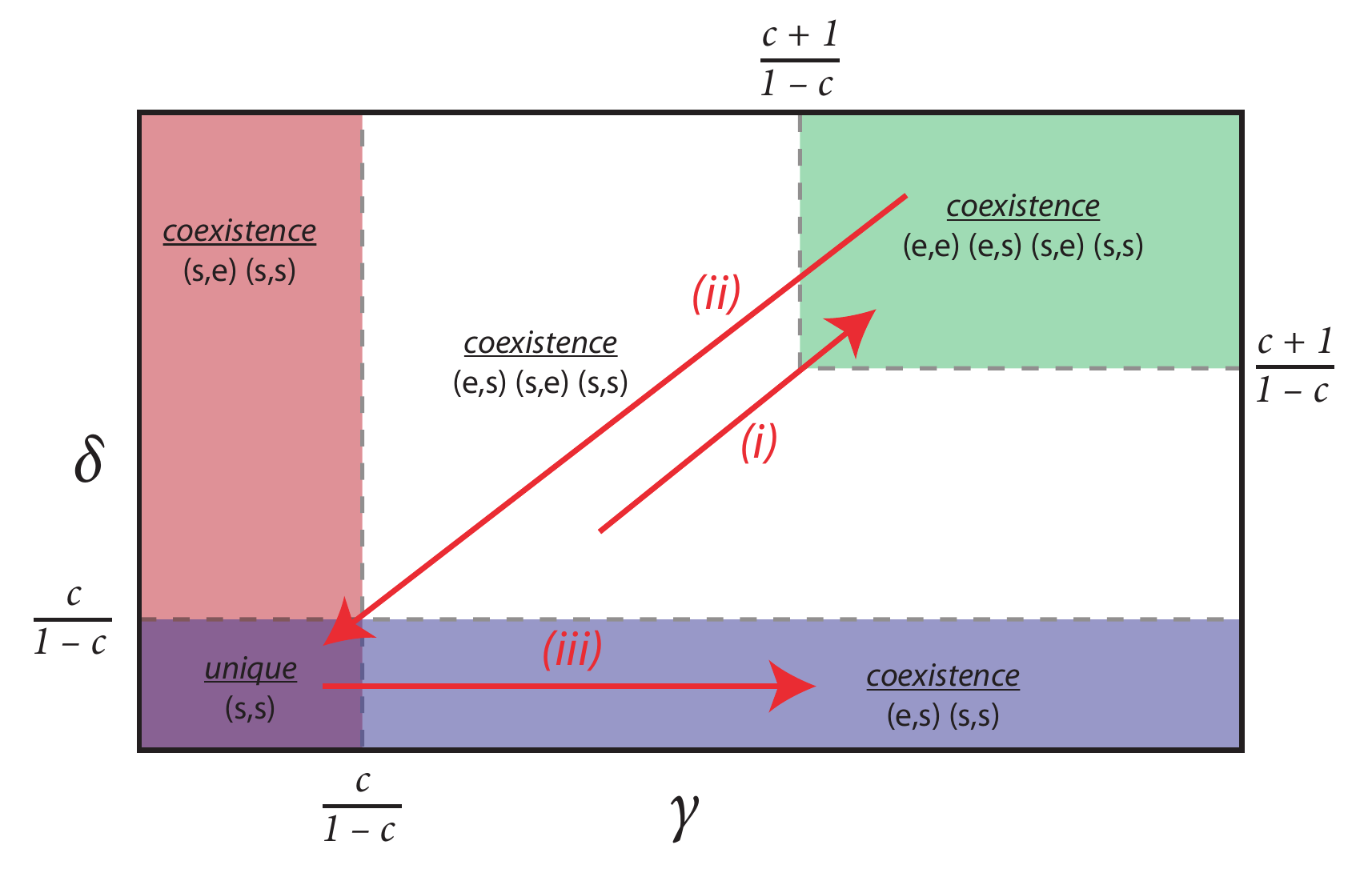}
    \caption{Illustration of the transitions between the game-theoretic equilibrium regions for (i) stronger internal cohesion of both opinion groups (`echo chambers'), (ii) less internal cohesion of both (heterophilious connections), and (iii) stronger internal cohesion for only one opinion group ('\#metoo').}
    \label{fig:metoo}
\end{figure}

\paragraph{Perception biases.}
In \cite{Sohngeidner}, the effect of the ego-network size, that is, the (average) number connections of the agents, on the occurrence of the spiral of silence was investigated. It was concluded that an increase in network density makes it more probable that one opinion group does not speak out publicly. In our work, we show that more density might even have the opposite effect. It depends on \textit{where} the additional connections are made: If new connections are guided by homophily, such that the opinion blocks become more cohesive, the spiral of silence might even be overcome (see path (i) in Figure \ref{fig:metoo}). We then arrive at a structure similar to `echo chambers,' in which only the voices affirming one's own view are heard and the others are blocked out (see \cite{baumann} for a contribution linking opinion dynamics to the emergence of echo chambers). If the additional connections are made between the opinion blocks, both $\gamma$ and $\delta$ decrease, which might make it more probable that the individuals have a more realistic picture of the overall opinion landscape. Then, the spiral effect is indeed more probable. But if the cross-group connections grow even further, both opinion groups misjudge their proportion to their own disadvantage, such that no group speaks out if there are costs associated to opinion expression (path (ii) in Figure \ref{fig:metoo}). The structure of the social contacts alone is already sufficient to cause misjudgements about opinion proportions in a social system. This is closely linked to more general accounts of perception biases \cite{lee2019homophily}.

\paragraph{Critical mass.} Furthermore, the model links to studies dealing with tipping points in social systems and the necessary numerical allocations, depending on the network structures. This has e.g. been analyzed for social conventions \cite{Centola}. 
If the social network of individuals is structured in opinion blocks, there is a hard numerical limit for the overcoming of a state in which one opinion is dominating publicly. For example, for a cross-opinion connection probability of $0.2$ (as in Figure \ref{fig:my_label}), the state in which both opinion camps are expressive cannot be reached if the minority makes up less than 22\% of the population.
\paragraph{Limits and outlook.} While we have stressed the generality of this work, we want to emphasize its limits as well: The homogeneous network structure of opinion blocks is not particularly realistic. Real social networks are rather heterogeneous, with well-connected and very active hubs and more `remote' individuals. Nevertheless, weighted (or stochastic \cite{wasserman1994social}) blocks can serve as a baseline for mathematical accessibility.

Moreover, this work is concerned with one way of reacting on social feedback, namely, the change in willingness to express one's opinion. Change in opinion is not included. It is probable that these phenomena take place on different time scales. Also, the social environments prompting opinion change might be different from the ones in which opinion predominance is fought for. In demonstrations, if two opinion camps meet each other, the main objective might not be information exchange or the need to convince each other, but to gain public audibility. Hence, a combination of models of opinion change and opinion expression might be in order in a multi-layer network approach, in which opinion formation and the competition for public opinion predominance take place on possibly different but interdependent network structures. \\ While there are plenty of studies on experimental evidence for the micromechanisms grounding the spiral of silence (see \cite{matthes} for a review), we are also seeking a more systematic larger-scale view on collective phenomena of opinion expression, which are closely related to the parameters $\gamma$ and $\delta$ in the model. A very prominent example of emerging collective opinion expression online, for which this model provides an explanation, is the Twitter-hashtag `\#metoo' and the subsequent movement against sexual harassment and sexual assault: Women found a device (in this case, a hashtag) that allowed them to find and connect to people who had experienced the same, and also to people who supported them. And all of a sudden, it was easier for them to speak out (path (iii) in Figure \ref{fig:metoo}). Measurements are an intricate task here: The networks one constructs out of interactions between individuals are only the networks \textit{of interaction}, that is, of only one part of the actions one wants to observe. Silent individuals do usually not show up in such networks since they are not involved in an observable way.

In conclusion, we develop a model of opinion expression which allows the investigation of how social structures can prevent or promote public opinion expression of different opinion groups. This approach allows direct connection to an influential theory of the social sciences, the spiral of silence \cite{NoelleNeumann1974,noelle2004spiral}. We approach the model both from a game-theoretic and from a dynamical systems perspective and show how the public audibility of certain opinions depends on the sensitivity of the agents towards their current evaluation of expected reward, the structural cohesion of the opinion groups and the costs for opinion expression.

\begin{acknowledgments}
This project has received funding from the European Union’s Horizon 2020 research and innovation programme under grant agreement No. 732942 (www.\textsc{Odycceus}.eu). We are grateful for the repeated discussion of the ideas described in the paper with Wolfram Barfuss, Sharwin Rezagholi, Florentin Münch, Armin Pournaki, and with the members of \textsc{Odycceus}. \\ \end{acknowledgments}

\appendix
\section{Expected decrease of the difference in $Q$-values}
We carry out the estimation for opinion group $G_{1}$, but the analogue holds for opinion group $G_{2}$. We can give an upper bound for the change in $Q$-value for the agent with the maximum $Q$-value of the group, $\dot{Q}_{i\in G_{1}}^{\text{max}}$, and a lower bound for the change in $Q$-value for the agent with the minimum $Q$-value of the group, $\dot{Q}_{i\in G_{1}}^{\text{min}}$ due to the monotonicity of the function $\frac{1}{1+e^{-x}}$:
\begin{widetext}
\begin{eqnarray}
\dot{Q}_{i\in G_{1}}^{\text{max}}=&&\alpha'(\frac{\gamma}{\gamma+1}\frac{1}{N_{1}-1}\sum_{\substack{j \in G_{1}\\ j\neq i}}\frac{1}{1+e^{-\beta Q_{j}}}-
\frac{1}{\gamma+1}\frac{1}{N_{2}}\sum_{j \in G_{2}}\frac{1}{1+e^{-\beta Q_{j}}}-Q_{i\in G_{1}}^{\text{max}}-c) \leq  \nonumber \\ 
&&\alpha'(\frac{\gamma}{\gamma+1}\frac{1}{N_{1}}(\sum_{\substack{j \in G_{1}\\ j\neq i}}\frac{1}{1+e^{-\beta Q_{j}}}+\frac{1}{1+e^{-\beta Q_{i\in G_{1}}^{\text{max}}}})
-\frac{1}{\gamma+1}\frac{1}{N_{2}}\sum_{j \in G_{2}}\frac{1}{1+e^{-\beta Q_{j}}}-Q_{i\in G_{1}}^{\text{max}}-c), \label{eq:ineq1}
\end{eqnarray}
\begin{eqnarray}
\dot{Q}_{i\in G_{1}}^{\text{min}}=&&\alpha'(\frac{\gamma}{\gamma+1}\frac{1}{N_{1}-1}\sum_{\substack{j \in G_{1}\\ j\neq i}}\frac{1}{1+e^{-\beta Q_{j}}}-
\frac{1}{\gamma+1}\frac{1}{N_{2}}\sum_{j \in G_{2}}\frac{1}{1+e^{-\beta Q_{j}}}-Q_{i\in G_{1}}^{\text{min}}-c) \geq \nonumber \\ 
&&\alpha'(\frac{\gamma}{\gamma+1}\frac{1}{N_{1}}(\sum_{\substack{j \in G_{1}\\ j \neq i}}\frac{1}{1+e^{-\beta Q_{j}}}+\frac{1}{1+e^{-\beta Q_{i\in G_{1}}^{\text{min}}}})
-\frac{1}{\gamma+1}\frac{1}{N_{2}}\sum_{j \in G_{2}}\frac{1}{1+e^{-\beta Q_{j}}}-Q_{i\in G_{1}}^{\text{min}}-c). \label{eq:ineq2}
\end{eqnarray}
\end{widetext}

If we now look at the change in time in the difference of $Q_{i\in N_{1}}^{\text{max}}$ and $Q_{i\in N_{1}}^{\text{min}}$, we can conclude by the above inequalities that the difference decreases at least exponentially in expectation by substracting the right hand-sides of (\ref{eq:ineq1}) and (\ref{eq:ineq2}).
\begin{eqnarray}
\frac{d}{dt}(Q_{i\in G_{1}}^{\text{max}}-Q_{i\in G_{1}}^{\text{min}}) \leq -\alpha' (Q_{i\in G_{1}}^{\text{max}}-Q_{i\in G_{1}}^{\text{min}}).
\end{eqnarray}
The analogue holds for opinion group $G_{2}$: 
\begin{eqnarray}
\frac{d}{dt}(Q_{i\in G_{2}}^{\text{max}}-Q_{i\in G_{2}}^{\text{min}}) \leq -\alpha' (Q_{i\in G_{2}}^{\text{max}}-Q_{i\in G_{2}}^{\text{min}}).
\end{eqnarray}

\end{document}